\documentclass[sigplan,nonacm]{acmart}

\setcopyright{none}

\usepackage{tikz}
\usepackage{amsmath}
\usepackage{pifont}

\usepackage[OT1]{fontenc}
\usepackage{algorithm}
\usepackage{algorithmicx}
\usepackage[noend]{algpseudocode}
\usepackage{textcomp}
\usepackage{booktabs} 
\usepackage{comment}
\usepackage{wrapfig}
\usepackage{xcolor}
\usepackage{graphicx}
\usepackage{array}
\usepackage{multicol}
\usepackage{afterpage}
\usepackage{float}
\usepackage{listings}
\usepackage{multirow}
\usepackage{arydshln}
\usepackage{xspace}
\usepackage{enumitem}
\usepackage{fancyhdr}
\usepackage{amsmath}

\usepackage{amssymb}
\usepackage{url}
\usepackage{todonotes}
\usepackage{dirtree}

\usepackage{tikz}
\usetikzlibrary{tikzmark}

\definecolor{FadedBanana}{RGB}{255,255,191}
\usepackage{makecell}

\usepackage[utf8]{inputenc}
\usepackage{newunicodechar}
\usepackage{lineno}

\usepackage{xurl}

\def\BibTeX{{\rm B\kern-.05em{\sc i\kern-.025em b}\kern-.08em T\kern-.1667em\lower.7ex\hbox{E}\kern-.125emX}}

\makeatletter
\renewcommand\section{\def\@toclevel{1}%
  \@startsection{section}{1}{\z@}%
  {6pt plus 4pt minus 2pt}%
  {4pt plus 2pt minus 2pt}%
  {\ACM@NRadjust\@secfont}}
\renewcommand\subsection{\def\@toclevel{2}%
  \@startsection{subsection}{2}{\z@}%
  {6pt plus 3pt minus 2pt}%
  {4pt plus 2pt minus 2pt}%
  {\ACM@NRadjust\@subsecfont}}
\renewcommand\subsubsection{\def\@toclevel{3}%
  \@startsection{subsubsection}{3}{\z@}%
  {6pt plus 3pt minus 2pt}%
  {4pt plus 2pt minus 2pt}%
  {\ACM@NRadjust{\@subsubsecfont\@adddotafter}}}
\renewcommand\paragraph{\def\@toclevel{4}%
  \@startsection{paragraph}{4}{\parindent}%
  {-.5\baselineskip \@plus -2\p@ \@minus -.2\p@}%
  {-3.5\p@}%
  {\ACM@NRadjust{\@parfont\@adddotafter}}}
\@ifundefined{ACM@origsection}{}{\let\ACM@origsection\section}
\@ifundefined{ACM@origsubsection}{}{\let\ACM@origsubsection\subsection}
\@ifundefined{ACM@origsubsubsection}{}{\let\ACM@origsubsubsection\subsubsection}
\@ifundefined{ACM@origparagraph}{}{\let\ACM@origparagraph\paragraph}
\makeatother
\usepackage{tikz}

\usepackage{cleveref}
\crefformat{chapter}{\S#2#1#3} 
\crefformat{section}{\S#2#1#3} 
\crefformat{subsection}{\S#2#1#3}
\crefformat{subsubsection}{\S#2#1#3}

\usepackage{subcaption}

\graphicspath{{figures/}}
\usepackage{pifont}

\newcommand{\phm}[1]{\vspace{.4em}
\noindent
\textbf{#1}\hspace{.5em}} 

\newcommand{\sys}{\textit{Mosaic}\xspace}

\newcommand{\set}[1]{\left \{ #1 \right \}}
\newcommand{\range}[1]{\left [ #1 \right ]}

\newcommand{\inner}[1]{\langle #1 \rangle}

\renewcommand{\@}[1]{\mathcal{#1}}

\newcommand{\op}[0]{\operatorname}

\newcommand{\Res}[0]{\@{G}}
\newcommand{\res}[0]{g}

\newcommand{\stage}[0]{S}
\newcommand{\Stage}[0]{\@S}

\newcommand{\alloc}{a}
\newcommand{\Alloc}{\@A}

\newcommand{\qsetmr}{\set{\alloc_m^\res}_{m\in S}^{\res \in \Res}}

\newcommand{\Tstage}{T_{\text{stage}}}
\newcommand{\Titer}{T_{\text{iteration}}}

\newcommand{\Tmodule}{T_{\text{module}}}

\begin{document}

\title{\sys: Towards Efficient Training of Multimodal Models with Spatial Resource Multiplexing
}
\author{Yanbo Wang}
\email{wang-yanbo@sjtu.edu.cn}
\affiliation{%
  \institution{Shanghai Jiao Tong University}
  \country{China}}
\affiliation{%
  \institution{Institute of Artificial Intelligence (TeleAI), China Telecom}
  \country{China}}

\author{Yuxuan Wang}
\email{fAKe@sjtu.edu.cn}
\affiliation{%
  \institution{Shanghai Jiao Tong University}
  \country{China}}

\author{Chen Chen}
\authornote{Corresponding author.}
\email{chen-chen@sjtu.edu.cn}
\affiliation{%
  \institution{Shanghai Jiao Tong University}
  \country{China}}

\author{Chunyu Xue}
\email{dicardo@sjtu.edu.cn}
\affiliation{%
  \institution{Shanghai Jiao Tong University}
  \country{China}}

\author{Yu Feng}
\email{y-feng@sjtu.edu.cn}
\affiliation{%
  \institution{Shanghai Jiao Tong University}
  \country{China}}

\author{Anbang Wu}
\email{anbang@cs.sjtu.edu.cn}
\affiliation{%
  \institution{Shanghai Jiao Tong University}
  \country{China}}

\author{Quan Chen}
\email{chen-quan@cs.sjtu.edu.cn}
\affiliation{%
  \institution{Shanghai Jiao Tong University}
  \country{China}}

\author{Yin Chen}
\email{cheny304@chinatelecom.cn}
\affiliation{%
  \institution{Institute of Artificial Intelligence (TeleAI), China Telecom}
  \country{China}}

\author{Qizhen Weng}
\email{wengqzh@chinatelecom.cn}
\affiliation{%
  \institution{Institute of Artificial Intelligence (TeleAI), China Telecom}
  \country{China}}

\renewcommand{\shortauthors}{Wang et al.}



\begin{abstract}

With the wide adoption of Multimodal Models (MMs) in real-world scenarios, it is significant to efficiently train the emerging MMs exhibiting increasingly complex module architectures. 
For MM deployment, existing works allocate a GPU to only one MM module following a temporal-multiplexing manner; this compromises the training efficiency because a single module often fails to attain high GPU utilization.
To improve GPU utilization and attain efficient MM training, we propose to deploy MMs in a temporal-spatial multiplexing manner, allowing multiple MM modules to colocate on a GPU with well-controlled resource quotas. 
In this paper, we propose \sys, an efficient MM training system applying temporal-spatial multiplexing. 
We first develop a flexible and lightweight execution engine that supports MM training with arbitrary resource quotas, and then build a comprehensive and accurate performance model to estimate module execution time under any possible allocation plan. 
With the performance model, we further adopt powerful heuristics to work out high-quality MM deployment plans in an efficient manner. 
Testbed experiments confirm that \sys can effectively improve the training efficiency of popular MMs, with a training speedup of up to 1.31$\times$.
\end{abstract}

\maketitle

\section{Introduction}
\label{sec:intro}

\begin{figure}[t]
    \vspace{.1in}
    \centering
    \begin{subfigure}{0.49\linewidth}
        \centering
        \includegraphics[scale=0.5]{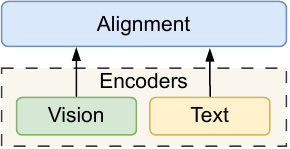}
        \caption{CLIP~\cite{clip}}
        \label{fig:subfig1}
    \end{subfigure}
    \hfill
    \begin{subfigure}{0.49\linewidth}
        \centering
        \includegraphics[scale=0.5]{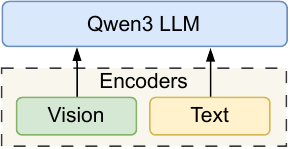}
        \caption{Qwen3-VL~\cite{qwen3vl}}
        \label{fig:subfig2}
    \end{subfigure}
    \caption{MMs comprise diverse dependent modules.}
    \label{fig:simple_mm}
\end{figure}

With the rapid development of AI techniques, \emph{Multimodal Models} (MMs)~\cite{internvl25, siglip2, llava, clip} are now widely adopted in many real-world scenarios, especially in edge devices for \emph{autonomous driving}~\cite{DriveVLM, DriveGPT4, DiMA} and \emph{embodied intelligence}~\cite{PaLM-E,RT-2,OpenVLA}.
As shown in Fig.~\ref{fig:simple_mm}, an MM is typically composed of multiple \emph{modules}: input encoders, a shared backbone, and output decoders, forming a directed acyclic graph.
In particular, given the deepening adoption of MMs in massive edge scenarios, an emerging trend for MM development is to maintain a modest model size (typical edge-grade MMs are smaller than 10B~\cite{MiniCPM-V, OmniVLM, OpenVLA, Megrez-Omni}) yet increase the number of involved modalities (i.e., from two or three modals to \emph{omni-modals}~\cite{ofasys,liu2025ola,team2026qwen3}).
Moreover, for competitiveness, such MMs often need to be routinely re-trained on the daily-collected data~\cite{FoMo-in-Flux, Modyn, MLLM-CL, TiC-CLIP}, rendering it of paramount significance to train them efficiently with the provisioned GPUs~\cite{distmm,spindle}. 

Given an MM with highly heterogeneous modules, it is challenging to deploy the MM modules over the GPU set for high efficiency.
For example, the classical \emph{Megatron-LM} scheme~\cite{megatron} parallelizes each module over all the GPUs,
which is inefficient due to the scaling overhead. 
\emph{DistMM}~\cite{distmm} and \emph{Spindle}~\cite{spindle} schemes 
instead partition the modules into execution \emph{waves} that temporally multiplex the cluster and, for each wave, minimize the cross-module execution time gap with module-adaptive GPU allocation. 
However, in those schemes each GPU is \emph{exclusively} allocated to \emph{only one module} at a time; in modern MMs with heterogeneous modules, some modules with low compute intensity often fail to achieve high utilization with their allocated GPUs, which further leads to compromised training efficiency (Fig.~\ref{fig:apollo_benefit_example}).
Therefore, to attain efficient MM training, we need to allow multiple modules to \emph{spatially multiplex} a GPU for improved utilization.
That is, overall we need to enforce \emph{temporal-spatial} multiplexing for MM deployment: 
first map the modules of an MM to sequentially-executed \emph{stages} for temporal GPU multiplexing, and then, within each stage, map each module to a set of compute SMs potentially on multiple GPUs. 
Ideally, with a proper \emph{MM-stage} and \emph{stage-GPU} mapping plan,
we can minimize the cross-module bubbles while attaining high GPU utilization, thereby optimizing the MM training efficiency. 
In realizing this insight, there are \emph{three key requirements}.
First, we need to maintain a flexible execution engine supporting module training with arbitrary SM allocations. 
Second, we need to build a comprehensive performance model to help evaluate any possible {stage-GPU} mapping solution.
Third, we need to find a high-quality mapping solution in an efficient manner.

In this paper, we propose \sys, an MM training system that exploits temporal-spatial GPU multiplexing for high efficiency. 
\sys is composed of three parts: (1) \emph{\sys Multiplexing Engine}---which allows multiple MM modules to arbitrarily multiplex a GPU in a flexible and lightweight manner, (2) \emph{\sys Performance Model}---which can accurately estimate the execution efficiency of a module under any possible spatial multiplexing plan, and (3) \emph{\sys Mapping Solver}---which can promptly find a high-quality temporal (MM-stage) and spatial (stage-GPU) multiplexing solution.

Specifically, we make non-trivial innovations in each solution aspect.
In the \emph{\sys Multiplexing Engine}, considering the SM partitioning granularity and isolation overheads, we choose to adopt the \emph{GreenContext} (GC) multiplexing technique~\cite{green_context}, which allocates fine-grained SM resources to a module with a separate GC-stream; we also propose \emph{stream-pool pre-creation} to mitigate GC preparation overheads in the critical path. 
Meanwhile, in the \emph{\sys Performance Model}, to support fractional SM allocation on multiple GPUs, we build a comprehensive \emph{scaling surface} for each module;
besides, to handle the performance interferences of colocated modules, we add both the \emph{additive} and \emph{multiplicative} rectification items 
to rectify the performance model, which 
proves to be more accurate than existing modeling methods. 
Finally, in the \emph{\sys Mapping Solver}, we jointly solve the \emph{MM-stage} and \emph{stage-GPU} mapping problem with high quality by combining the \emph{Greedy Agglomerative Hierarchical Clustering} (GAHC) heuristic with the Google CP-SAT solver; we also incorporate the \emph{early-pruning} and \emph{result-caching} techniques for even better solving efficiency.

We have implemented \sys with 8K LoCs, and evaluated its performance in a testbed with 32 H100 GPUs. 
Our end-to-end experiments with popular MMs show that, by making higher GPU utilization with spatial multiplexing, \sys improves the training efficiency by up to 1.31$\times$ over the state-of-the-art methods.
Meanwhile, our deep-dive experiments further confirm that each \sys innovation---in the multiplexing engine, performance model, as well as mapping solver---does make non-negligible contribution to the overall performance superiority. 


In summary, this paper makes three key contributions.
\begin{itemize}[leftmargin=*]
    \item We identify \emph{temporal-spatial GPU multiplexing} as an effective deployment paradigm for training heterogeneous MMs, where dependent modules temporally share GPUs and parallel modules spatially share SM resources.
    \item We design \sys, a practical MM training system that realizes this paradigm through a GreenContext-based multiplexing engine, an interference-aware performance model, and a joint MM-stage/stage-GPU mapping solver.
    \item We implement \sys and demonstrate on a 32-H100 testbed that it improves training efficiency by up to $1.31\times$ over state-of-the-art MM training schemes.
\end{itemize}

\section{Background and Motivation}
\label{sec:background}

\subsection{Multimodal Models: The Emerging Trend}
\label{sec:research_background}

\begin{figure}[t]
  \centering
  \includegraphics[width=0.85 \linewidth]{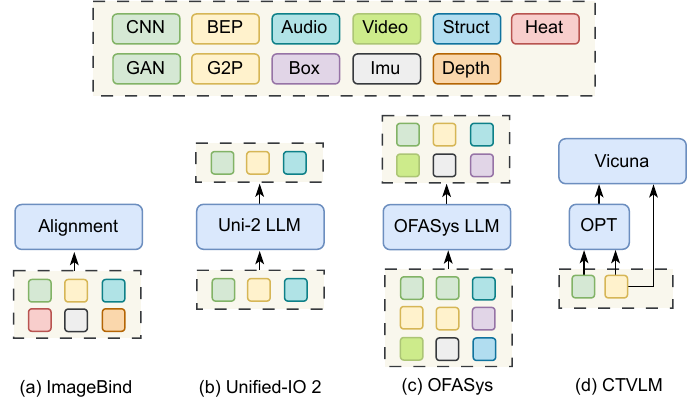}
    \caption{Edge-grade MMs with high modal complexity.} 
    \label{fig:complex_mm}
\end{figure}

Multimodal Models (MMs), which jointly process inputs of diverse forms (e.g., \emph{text}, \emph{vision}, and \emph{audio})~\cite{gpt-4,gemini25,llava}, are increasingly significant for diverse scenarios like \emph{embodied intelligence}~\cite{liu2026insight} and \emph{autonomous driving}~\cite{auto_driving}.
As illustrated in Fig.~\ref{fig:simple_mm} and Fig.~\ref{fig:complex_mm}, an MM is composed of multiple (encoder, decoder, and backbone) modules with mutual dependencies, which form a \emph{Directed Acyclic Graph} (DAG).
For example, \emph{CLIP}~\cite{clip} is a classical MM that aligns the features of different modalities via contrastive learning, and \emph{Qwen3-VL}~\cite{qwen3vl} uses an LLM backbone to merge the inputs respectively from the \emph{vision} and \emph{text} encoders.

In particular, given the widespread and also in-depth adoption of MMs in many real-world application scenarios~\cite{sirusurvey,edge_ai_market,Multimodal_market}---where the MMs are increasingly deployed on \emph{edge hardware} like cars, robots, and phones~\cite{drivelm,threedvla,minicpmv_natcomm,imp,OpenVLA}, the recent days have witnessed a \emph{trend} to develop \emph{modest-size} yet \emph{high-modal-complexity} MMs: maintaining a modest size allows the MM to be hosted by low-end edge devices, and supporting more modalities can make the MM more powerful for realistic tasks. 
Fig.~\ref{fig:complex_mm} shows such MM examples. 
\emph{ImageBind}~\cite{imagebind} aligns 6 modalities (\emph{image}, \emph{text}, \emph{audio}, \emph{depth}, \emph{thermal} and \emph{IMU}) into one embedding space, \emph{Unified-IO~2}~\cite{unifiedio2} combines 5 encoders and 3 decoders with an LLM backbone, and \emph{OFASys}~\cite{ofasys} integrates 9 encoders and 6 decoders with a universal model. 
As revealed by Table~\ref{tab:lmm_module_config}, those modules have strong architecture heterogeneity.
Moreover, as shown in Fig.~\ref{fig:complex_mm}d, in some cases multiple MMs may work together on a shared device~\cite{liu2026insight,wu2025multi}, which also form a \emph{logically-complex} MM requiring joint training.

Meanwhile, for competitiveness, MMs need to be timely refreshed with routinely collected data. 
Recent studies already report treating MM updating as daily workloads: FoMo-in-Flux~\cite{FoMo-in-Flux} and D-MoLEs~\cite{dmole} have both focused on \emph{continual MM training} in real-world scenarios,
and Modyn~\cite{Modyn} has further explored how to orchestrate \emph{recurring} model updates in data-centric pipelines. 
Hence, it is \emph{increasingly significant} to train MMs \emph{efficiently} with the provisioned GPU set.


\begin{table}[t]
\centering
\setlength{\tabcolsep}{1.2pt}      
\renewcommand{\arraystretch}{0.9} 
\small                           

\begin{tabular*}{\linewidth}{@{\extracolsep{\fill}}cccccc} 
\toprule
\textbf{Model} & \textbf{Module} & \textbf{Layers} & \textbf{Dim.} & \textbf{TFLOPs} & \textbf{CI} \\
\midrule
\multirow{3}{*}{\makecell{Qwen3-VL\\(8.1B)}}
      & Qwen3LLM     & 36 & 4096 & 22.27 & 145.2 \\
      & Vision       & 27 & 4096 & 2.58  & 82.4  \\
      & Text         & 1  & 4096 & 0.15  & 2.1   \\
\midrule
\multirow{4}{*}{\makecell{Unified-IO 2\\(3.8B)}}
      & UIO-2 LLM    & 48 & 3072 & 16.70 & 110.5 \\
      & Vision       & 11 & 768  & 1.48  & 24.6  \\
      & Audio        & 11 & 768  & 1.06  & 21.8  \\
      & Text         & 1  & 3072 & 0.10  & 4.5   \\
\midrule
\multirow{3}{*}{\makecell{ImageBind\\(1.2B)}}
      & Vision       & 24 & 1024 & 4.17  & 35.2  \\
      & Audio        & 12 & 768  & 2.09  & 22.8  \\
      & Text         & 12 & 768  & 1.04  & 20.5  \\
\midrule
\multirow{4}{*}{\makecell{OFASys\\(6.3B)}}
      & OFASys LLM   & 36 & 1280 & 4.80  & 41.6  \\
      & Vision       & 8  & 1280 & 1.35  & 18.2  \\
      & Text         & 4  & 1280 & 0.72  & 12.5  \\
      & Audio        & 6  & 1280 & 0.95  & 14.8  \\
\bottomrule
\end{tabular*}
\caption{Architecture information of representative MMs (for TFLOPs information, the input configuration is the same as in Table~\ref{tab:mockdata_config}). \emph{Compute intensity} (CI) is measured in FLOPs/Byte.}
\label{tab:lmm_module_config}
\end{table}

\subsection{Efficiency-oriented MM Deployment}
\label{subsec:existing_solution}

Optimizing model deployment for high training efficiency is a classical research problem in the literature, yet it remains challenging for MMs comprising heterogeneous modules organized in DAGs.
Essentially, deploying a DAG-structured MM involves two decision aspects: (1) how to deploy the \emph{dependent} modules, and (2) how to deploy the \emph{parallel} modules. 
Here, we walk through existing deployment methods concerning the two aspects. 

\phm{Deploying dependent MM modules.}
Regarding the deployment of dependent MM modules, the data dependencies between upstream and downstream modules prevent their concurrent execution.
For conventional \emph{Large Language Models} (LLMs) typically of a large size, the dependent modules (aka model layers) are often deployed in GPU clusters following the \emph{Pipeline Parallel} (PP) paradigm~\cite{pipedream, gpipe}. 
However, PP deployment incurs \emph{warm-up} and \emph{cool-down} bubbles which waste the GPU resources; for our targeting \emph{edge-grade MMs} which are not large (and can thus be kept in GPU memory during training), it would be more efficient to have them \emph{temporally-multiplex} the GPUs~\cite{megatron,distmm,spindle}, such that no GPUs have to stand by idly waiting for the output of upstream modules. 
Then we turn to deploying the \emph{parallel} modules, which are particularly challenging for MMs due to the substantial module heterogeneity.

\begin{figure}[t]
    \centering
    \begin{subfigure}[t]{0.31\linewidth}
        \centering
        \makebox[\linewidth][c]{%
            \includegraphics[height=2.5cm,keepaspectratio]{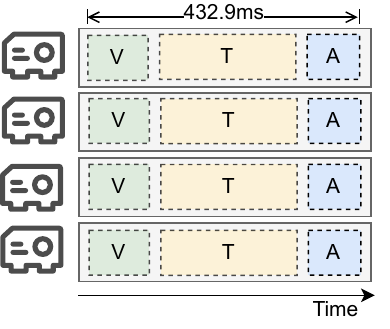}
        }
        \caption{Megatron-LM}
        \label{fig:megatron}
    \end{subfigure}
    \hspace{0.01\linewidth}
    \begin{subfigure}[t]{0.30\linewidth}
        \centering
        \makebox[\linewidth][c]{%
            \includegraphics[height=2.5cm,keepaspectratio]{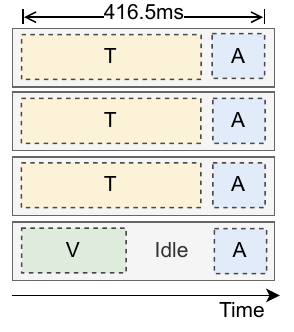}
        }
        \caption{DistMM}
        \label{fig:distmm}
    \end{subfigure}
    \begin{subfigure}[t]{0.30\linewidth}
        \centering
        \makebox[\linewidth][c]{%
            \includegraphics[height=2.5cm,keepaspectratio]{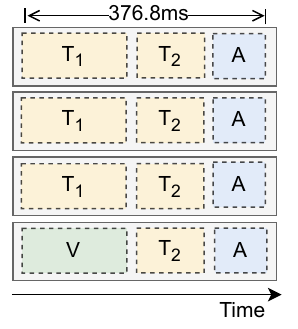}
        }
        \caption{Spindle}
        \label{fig:spindle}
    \end{subfigure}
    \caption{Behaviors of different MM deployment schemes when training the CLIP model on four GPUs. \textit{V}, \textit{T}, and \textit{A} denote \emph{vision-encoder}, \emph{text-encoder} and \emph{alignment} modules.
}
    \label{fig:existing_solution}
\end{figure}

\phm{Deploying parallel MM modules.}
Regarding the deployment of parallel (yet heterogeneous) MM modules, a series of works have been proposed in the literature.
As illustrated in Fig.~\ref{fig:megatron}, the classical \emph{Megatron-LM} framework~\cite{megatron} adopts a symmetric allocation strategy, copying each module to all the provisioned GPUs following the \emph{Data Parallel} (DP) paradigm~\cite{dp} (different modules are still executed sequentially via temporal multiplexing). 
Although straightforward, this paradigm is often inefficient: the usually over-aggressive parallelization would amplify the communication-to-computation ratio and incur remarkable per-module execution slowdown. 

To address such inefficiency, later MM deployment works enable \emph{asymmetric}, \emph{module-adaptive} GPU allocation.
As shown in Fig.~\ref{fig:distmm}, \emph{DistMM}~\cite{distmm} assigns disjoint sets of GPUs to different modules (each module still enforces DP over its allocated GPUs). 
Its objective is to balance the execution times of parallel modules; however, because resource allocation is restricted to integer GPU counts, the training system frequently suffers from sub-optimality due to rounding error. 
These errors may leave substantial duration misalignment between modules, resulting in resource idle time prior to the cross-module data-merging barriers.  
To further alleviate such rounding-caused module duration inconsistency,
as shown in Fig.~\ref{fig:spindle}, \emph{Spindle}~\cite{spindle} proposes to decompose the modules into finer-grained slices. 
Note that, although this improved granularity allows for better temporal alignment, it comes at the cost of significantly higher programming complexity and increased coordination overhead due to the frequent synchronization required between slices.

\phm{Common limitations of existing deployment paradigms.}
However, despite the differences, the aforementioned paradigms commonly exhibit a fundamental limitation: they assign each GPU to only one MM module, assuming that \emph{each module must exclusively occupy the GPUs allocated to it}; such an allocation restriction inherently compromises the resultant MM training efficiency.

In fact, it often occurs that a single MM module fails to attain high utilization on the GPUs allocated to it.
Specifically, in Table~\ref{tab:lmm_module_config}, we list the module-level architecture information of typical MMs, including their \emph{Compute Intensity} (CI).
Table~\ref{tab:lmm_module_config} suggests that the per-module compute intensity can vary by over \emph{an order of magnitude} across different modules, exhibiting strong cross-module heterogeneity.
For example, for the Qwen3-VL-8B model, the \emph{text-encoder} module has a CI of 2.1 FLOPs/Byte, the \emph{vision-encoder} module has a CI of 82.4 FLOPs/Byte, yet the LLM backbone as a module has a CI of 145.2 FLOPs/Byte. 
In the meantime, modern GPUs feature hundreds of \emph{streaming multiprocessors} (SMs)~\cite{sm}, and a module with low CI usually fails to fully monopolize those SMs, thereby yielding low GPU utilization.

\begin{figure}[t]
    \centering
    \captionsetup[subfigure]{justification=centering}

    \begin{subfigure}[t]{0.47\linewidth}
        \centering
        \vspace{0pt}
        \includegraphics[height=3.4cm,keepaspectratio]{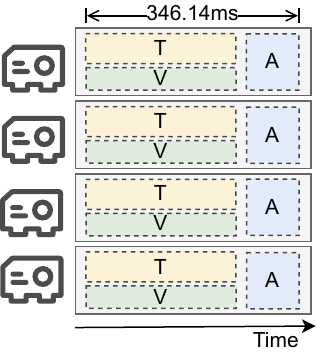}
        \caption{A Colocated Method}
        \label{fig:apollo_schedule}
    \end{subfigure}
    \hfill
    \begin{subfigure}[t]{0.52\linewidth}
        \centering
        \vspace{0pt}
        \includegraphics[height=3.4cm,keepaspectratio]{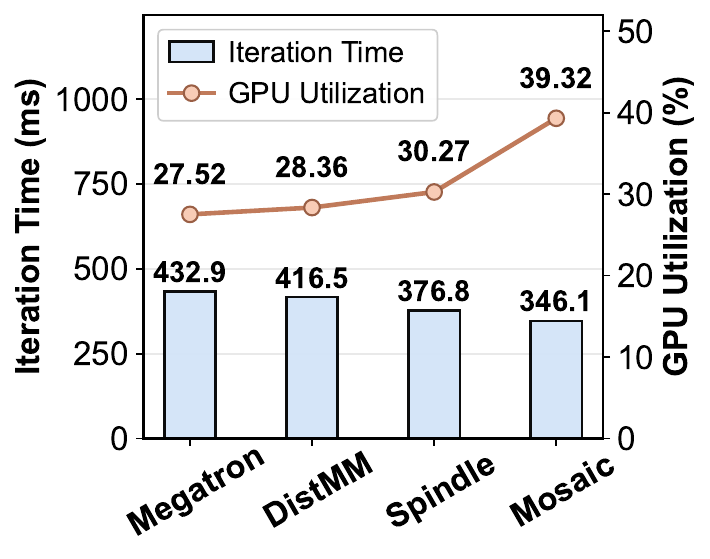}
        \caption{Efficiency/utilization Results}
        \label{fig:sys_compare}
    \end{subfigure}

    \caption{Benefits of spatial GPU multiplexing. Allowing vision (V) and text (T) encoders to colocate reduces total iteration time and improves overall GPU utilization.}
    \label{fig:apollo_benefit_example}
\end{figure}

By contrast, by relaxing the exclusive GPU allocation constraint, we can potentially attain higher GPU utilization and further yield faster MM training. 
As shown in Fig.~\ref{fig:apollo_schedule}, with modern spatial multiplexing techniques \cite{green_context,mps,mig}, we can colocate multiple MM modules on a GPU.
By properly allocating a portion of the GPU SMs to each module based on its computing characteristics, it is possible that each colocated module can complete at around the same time.
In this way, the provisioned GPUs can (1) avoid temporal resource wastage while (2) attaining higher utilization during their busy time.
To confirm, we train the CLIP model on a $4\times$H100 GPU server under different deployment paradigms.
As revealed by Fig.~\ref{fig:sys_compare}, compared to those existing paradigms in Fig.~\ref{fig:existing_solution}, a \emph{spatial-multiplexing-enabled} deployment solution illustrated in Fig.~\ref{fig:apollo_schedule} can indeed attain a higher GPU utilization (with an improvement of 29.9\% over the status quo \emph{Spindle} paradigm) and further a shorter iteration time.
This result confirms the necessity to enable spatial GPU multiplexing in MM deployment. 

\phm{Insight.}
In summary, when training MMs comprising heterogeneous modules, we need to exploit \emph{temporal-spatial GPU multiplexing} for best training efficiency. 
For \emph{dependent} MM modules, we need to adopt \emph{temporal} GPU multiplexing to avoid resource bubbles;
more importantly, for \emph{parallel} MM modules, we need to adopt \emph{spatial} GPU multiplexing to enhance the GPU utilization. 
Next, we will explore how to enforce this insight into practice.

\section{\sys Solution Design}
\label{sec:solution} 

In this section, we present the design details of \sys, an MM training system exploiting \emph{temporal-spatial GPU multiplexing} for high compute efficiency. 
We will first mathematically formulate the problem and make the system overview in Sec.~\ref{sec:problem_formulation}, and then we elaborate each design component in Sec.~\ref{subsec:apollo_engine}, Sec.~\ref{subsec:performance_modeling}, and Sec.~\ref{subsec:mapping_solver}. 
Finally, we discuss some peripheral design issues in Sec.~\ref{sec:discussion}. 


\subsection{Problem Formulation and System Overview} \label{sec:problem_formulation} 

\phm{Problem formulation.} 
For clarity, we first formulate the problem of efficiency-oriented MM deployment.
We represent each MM training job as a DAG, $\inner{\@M, \@E}$, where $\@M$ denotes the MM \emph{modules} and $\@E$ denotes the data \emph{dependencies} between modules.
Under the temporal-spatial multiplexing mode, MM training needs to be executed in sequential \emph{stages}, each comprising a set of parallel modules: modules in different stages temporally multiplex the GPU resources, whereas modules within a stage spatially multiplex the GPU resources.
Following this mode, the MM deployment problem can be expressed as conducting \emph{two-level} mappings: (1) \emph{MM-stage mapping}---map the MM modules to temporally ordered stages with the module dependency preserved, and (2) \emph{stage-GPU mapping}---map the parallel modules of each stage to the GPU resources with fine-grained SM allocation enabled.
Next, we respectively formulate the two mapping problems.

\textit{1) Formulating the outer \textit{MM-stage} mapping problem.}
Suppose the execution plan consists of $n$ ($n$ is also a decision variable) stages $\Stage = \inner{\stage_1, \stage_2, \ldots, \stage_n}$, which are executed sequentially on a pool of GPU resources $\Res$. 
We let $\Titer$ represent the training iteration time which is the optimization target, 
then the MM-stage optimization problem can be formulated as:

\begin{align}
    \Titer^*(\@M, \Res) =& \min_{n,\Stage=\inner{\stage_1, \stage_2, \ldots, \stage_n}} \sum_{i=1}^{n}  \Tstage^*(\stage_i, \Res), \label{eq:inter_stage_obj}\\
     \text{s.t.}\quad & \bigcup_{i\in[n]} \stage_i = \@M, \label{eq:inter_stage_cons1} \\
                 & \Stage^{-1}(m) < \Stage^{-1}(m'), \text{\quad} \forall (m, m') \in \@E. \label{eq:inter_stage_cons2}
\end{align}

The objective formula \eqref{eq:inter_stage_obj} means to minimize total iteration time, which, given the temporal multiplexing nature, is modeled as the sum of all stages' execution times.
Here $\Tstage^*(\stage_i, \Res)$ means the shortest possible execution time of stage $\stage_i$ given the GPU set $\Res$, which is obtained by solving the later stage-GPU mapping problem.
The constraint formula \eqref{eq:inter_stage_cons1} states that all modules are covered, and formula set \eqref{eq:inter_stage_cons2} ensures the data dependency is complied with by the mapping plan, where $\Stage^{-1}(m)$ denotes the index of the stage that contains module $m$. 

\textit{2) Formulating the inner \textit{stage-GPU} mapping problem.}
In solving the above optimization problem, for each candidate MM-stage mapping solution (i.e., fixing $\stage_i$ and $\Res$), 
we need to acquire $\Tstage^*(\stage_i,\Res)$---the shortest stage execution time attained under the optimal module deployment (stage-GPU mapping) plan.
Specifically, we let $\alloc_{m}^{\res} \in [0, 1]$ be the SM quota of GPU $\res$ allocated to module $m$, $\Alloc=\qsetmr$ be the overall stage-GPU mapping plan,
and $\Tmodule(m, \Alloc)$ be the execution time of module $m$.
Then the stage-GPU mapping problem can be formulated as:



\begin{align}
    \Tstage^*(\stage,\Res)=& \min_{\Alloc=\qsetmr} \max_{m\in \stage} \Tmodule(m, \Alloc) \label{eq:stage_time} \\
    \text{s.t.}\quad &\sum_{m \in \stage} \alloc_m^\res \le 1, \text{\qquad}  \forall \res \in \Res.  \label{eq:comp_lim}
\end{align}

The objective formula \eqref{eq:stage_time} aims to minimize the end-to-end stage latency, estimated as the slowest module's latency.
Constraint formulas \eqref{eq:comp_lim} prevent over-allocation of computation resources. 
The memory resource constraint can be formulated similarly; it is omitted here for simplicity, but always complied with in practice.
Note that existing methods like \emph{DistMM} and \emph{Spindle} essentially restrict the allocation $\alloc_m^\res$ to binary values $\set{0, 1}$, and from a mathematical point of view, such a solution space restriction naturally compromises the solution quality.

\phm{Challenges.}
While the above formulas clearly depict what an ideal solution should be, 
it is still difficult to find and apply that optimal solution in practice. 
Specifically, there are three challenges. 
\begin{itemize}
    \item First, to support spatial multiplexing with controlled GPU quota, we need to prepare a flexible and light-weight training engine that can execute MM modules with arbitrary SM allocation amounts. 
        While modern GPUs support a series of spatial multiplexing techniques~\cite{mps,mig,green_context}, they, however, lack the allocation amount flexibility or bring large overheads when preparing the desired SM execution slots for MM modules in each stage.
    \item Second, to solve the above optimization problems, a prerequisite is to obtain a comprehensive performance model, $\Tmodule(m, \Alloc)$, that can estimate the execution time of a module when colocated with other modules at arbitrary resource ratios.
        However, with partial GPU allocation allowed, the performance modeling space is much larger than existing ones; worse, due to complicated interferences among colocated modules, it is hard to accurately predict the module performance for arbitrary colocating plans.
    \item Third, it also presents a non-trivial challenge to solve the formulated problems in a practical manner. 
    Given the increasing MM complexity and the nested nature of the two optimization problems, the solution space is highly complex, and it is hard to find a high-quality deployment solution within a modest time budget. 
\end{itemize}
To harvest the benefit of temporal-spatial multiplexing in real-world MM deployments, we need to address all three challenges in our proposed system.

\begin{figure}[t]
  \centering
  \includegraphics[width=0.9 \linewidth]{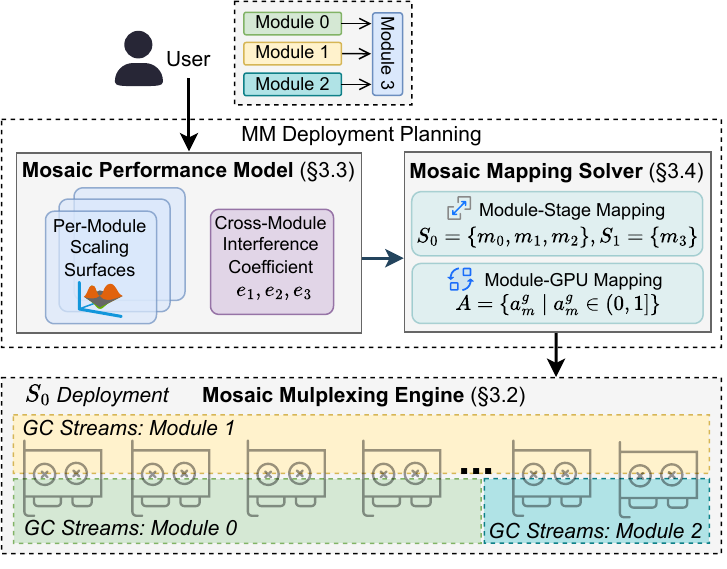}
  \caption{Overview of our \sys system.}
  \label{fig:overview}
\end{figure}

\phm{System overview.}
As shown in Fig.~\ref{fig:overview}, we develop \sys, an MM training system that exploits temporal-spatial GPU multiplexing for high training efficiency.
\sys is composed of three components: the \emph{\sys Multiplexing Engine}, the \emph{\sys Performance Model}, and the \emph{\sys Mapping Solver}. 

\begin{itemize}
    \item First, the \emph{\sys Multiplexing Engine} supports module training with an arbitrary number of SMs on a GPU.  To support such flexibility while being light-weight for runtime use, we adopt the \emph{Green Context} (GC) technique and further propose \emph{GC-stream pool pre-creation} to amortize the maintenance overheads.  
    \item Second, the \emph{\sys Performance Model} can accurately estimate the execution time of an MM module when it is allocated any number of GPU SMs and colocated with any MM modules. 
    We build a \emph{scaling surface} for comprehensive modeling and also incorporate the \emph{additive} and \emph{multiplicative} interference items for accurate modeling.
    \item Third, the \emph{\sys Mapping Solver} can efficiently find a high-quality temporal (MM-stage) and spatial (stage-GPU) multiplexing plan for an MM.
It combines the \emph{Greedy Agglomerative Hierarchical Clustering} (GAHC) heuristic with the \emph{CP-SAT} solver to get high-quality solutions, and also adopts \emph{early-pruning} and \emph{result-caching} techniques to further enhance the solving efficiency. 
\end{itemize}

Next, we elaborate on the solution design in each \sys component in greater detail.

\subsection{Flexible and Light-weight Spatial Multiplexing}
\label{subsec:apollo_engine}

To exploit spatial multiplexing for efficient MM training, we need to first provide a flexible training runtime.
Specifically, that runtime should (1) support \emph{arbitrary partitioning} of a GPU's computing resources to different MM modules---per the solution requirement of Eq.~\ref{eq:stage_time}. 
Moreover, as indicated in Fig.~\ref{fig:apollo_schedule}, that GPU partitioning scheme should (2) be \emph{dynamically-tunable} during the training process---given the periodical occurrence of stage transitions. 

\begin{figure}[t]
  \centering
  \begin{subfigure}[t]{0.48\linewidth}
    \centering
    \includegraphics[width=1\linewidth, height=3.8cm, keepaspectratio]{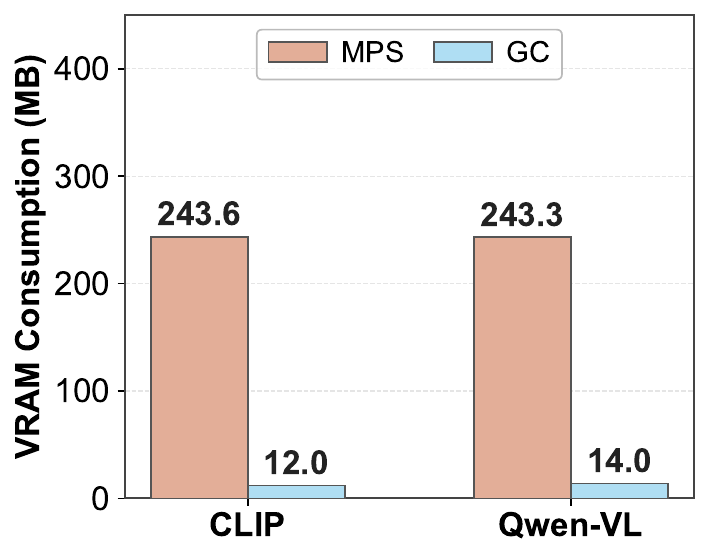}
    \caption{VRAM overhead comparison between MPS and Green Context. Green Context features 95\% fewer memory overhead.}
    \label{fig:memory}
  \end{subfigure}
  \hfill
  \begin{subfigure}[t]{0.48\linewidth}
    \centering
    \includegraphics[width=1\linewidth, height=3.8cm, keepaspectratio]{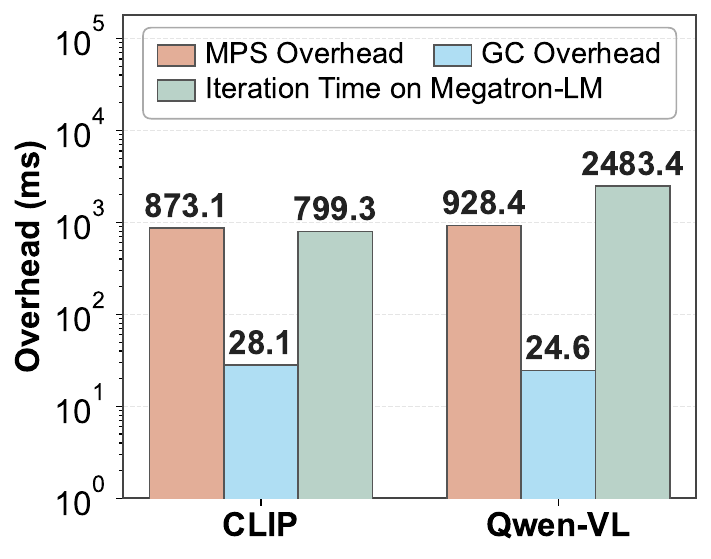}
    \caption{Latency overhead comparison. Green Context achieves the least overhead, with a latency of less than 30 ms.}
    \label{fig:latency_overhead}
  \end{subfigure}

  

  \caption{Green Context is lightweight in both memory and time overhead.}
  \label{fig:gctx_analysis}
\end{figure}

\phm{Choosing among popular multiplexing techniques for best flexibility.} 
Based on the two requirements above, we surveyed the popular spatial multiplexing techniques supported by modern NVIDIA GPUs. 
The Multi-Instance GPU (MIG)~\cite{mig} technique can partition a 
GPU into up to seven instances, each fully isolated with its own memory, cache, and compute cores. 
However, the MIG technique cannot partition GPU SMs at arbitrary portions (must follow fixed configurations). 
Meanwhile, Multi-Process Service (MPS)~\cite{mps} is a classical spatial multiplexing technique that allows allocating an arbitrary number of GPUs
to a \emph{process} (i.e., a MM training job in our problem). 
However, MPS itself does not support fine-grained, intra-process resource isolation at per-module level.
Recently, \emph{Green Context} (GC)~\cite{green_context} has been developed as an advanced spatial multiplexing technique that supports intra-process SM isolation at CUDA stream level; besides, as shown in Fig.~\ref{fig:gctx_analysis}, its memory and time overheads are much less than MPS. 
Therefore, we choose to adopt the GC techniques in \sys: by mapping a MM module to a CUDA-stream with designated SM quota, we can realize module-level adaptive GPU resource allocation. 




\phm{Light-weight allocation adjustment with a pre-created stream pool.}
While Green Context supports fine-grained GPU multiplexing, once a GC stream is launched, its SM quota can however not be adjusted at runtime.
Therefore, we need to prepare a distinct Green Context stream (GC-stream) for each module with the desired SM quota. 
Nonetheless, creating or destroying such GC-stream would incur non-negligible overheads in the critical path (e.g., for reclaiming the stream objects as well as the associated GC states). 
Our later testbed evaluations (Fig.~\ref{fig:gctx_controller_speedup}) show that, when training the Imagebind model on 8 × H100, it takes up to 6.63\% of the per-iteration time to manage the GC-streams in a on-demand (create-and-then-destroy) manner. 
Such overheads would be substantial when conducting stage transition for multiple times within each MM training iteration.

Therefore, to eliminate stream preparation overheads in the critical path, we choose to create a Green Context \emph{stream pool} in advance.
Moreover, we note that MM training is an iterative process where the same spatial-colocation plan would be repeated in each iteration.
That is, given the MM modules and the available GPUs, all the desirable GC-stream configurations to appear is essentially a white-box information---acquirable with offline profiling. 
Therefore, in our \sys system, we pre-create---at training commencement---a pool of GC-streams-to-use with the desired SM quotas; once a training stage completes in an iteration, the modules of the next training stage can be immediately launched, without incurring any stream-creation/destruction delay in the critical path. 
This way, we can make flexible and light-weight resource provisioning at per-module granularity, well supporting the enforcement of our solution paradigm.

\subsection{Thorough and Accurate Performance Modeling}
\label{subsec:performance_modeling}


With spatial sharing enabled, a MM module can be allocated an arbitrary SM portion on each GPU: $\alloc_{m}^{\res} \in \range{0, 1}$. This, however, complicates the estimation of $\Tmodule(m, \Alloc)$ (the latency of module $m$ under allocation $\Alloc$) for two reasons: (1) The allocation space expands from a discrete set of GPU counts to a continuous multi-dimensional surface, making exhaustive profiling infeasible; (2) Spatial sharing introduces complex interference patterns when modules are collocated onto the same GPU. These challenges render existing solutions inadequate and here we respectively address them.

\phm{Towards comprehensive modeling with \emph{symmetry-based solution pruning} and \emph{smoothness-based grid sampling}.} 
To overcome the first challenge, \sys builds a \emph{scaling surface} through symmetry-based pruning and smoothness-based sampling.
While \sys theoretically allows a module to receive heterogeneous SM quotas across different GPUs, such asymmetric configurations are inherently suboptimal for data-parallel replicas: synchronous iterations are limited by the slowest worker, any asymmetry creates stragglers and results in wasted resource cycles on faster GPUs. Consequently, we prune the search space by enforcing symmetric allocations. Let $\Res_m = \{\res \in \Res \mid \alloc_m^\res > 0\}$ be the set of GPUs assigned to module $m$, and $d_m = |\Res_m|$ be its data-parallel degree, we then require that:
\begin{equation}
  \forall m, \exists \alloc_{m}, \forall \res \in \Res_m, \quad \alloc_m^\res = \alloc_{m}.
  \label{eq:symmetric_quota_pruning}
\end{equation}
By enforcing a uniform quota $\alloc_m$ across all $d_m$ replicas, we effectively collapse the high-dimensional allocation mesh into a tractable two-dimensional space -- each module's deployment is thus simplified to a tuple $(d_m, \alloc_m)$. 

\begin{figure}[t]
  \centering
  \includegraphics[width=0.9 \linewidth]{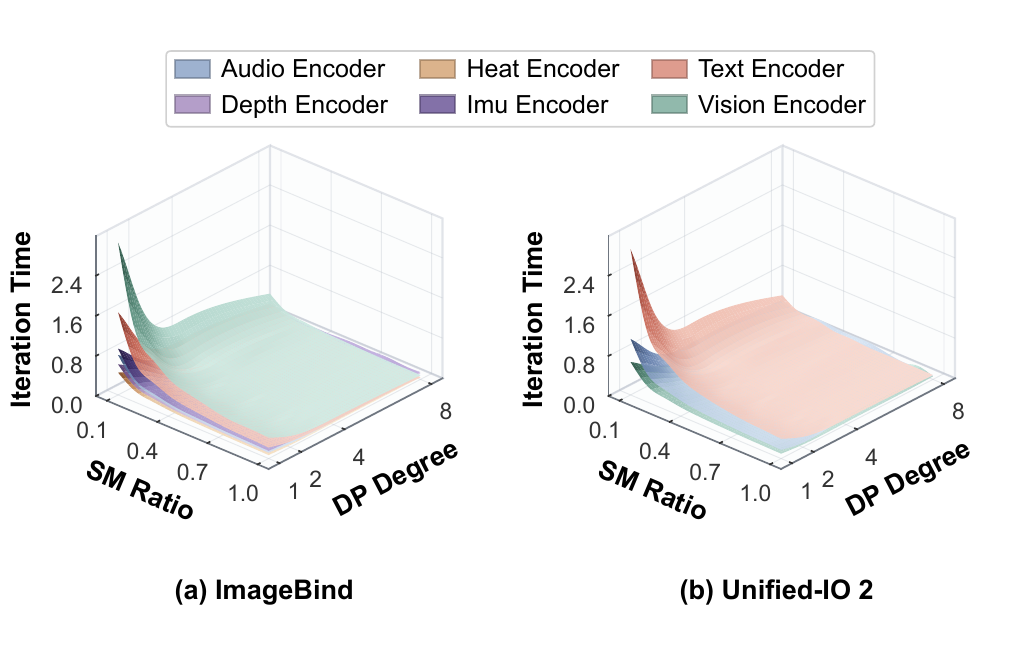}
  \caption{MM modules' scaling curves are smooth with respect to DP Degree and SM Ratio.}
  \label{fig:sm_performance_model}
\end{figure}

Since empirical observations in Fig.~\ref{fig:sm_performance_model} show that latency varies smoothly along both dimensions, \sys approximates this surface with sparse grid sampling rather than exhaustive measurement. Specifically, we profile $d_m$ at powers-of-two (e.g., $1, 2, 4, 8$) and $\alloc_m$ at \emph{decile} increments (i.e., $0.1, 0.2, \dots, 1.0$). This sparse mesh provides a high-fidelity approximation of the scaling surface $\Tmodule(m, \Alloc) = f_m^{\text{scaling surface}}(d_m, \alloc_m)$ while reducing profiling overhead by orders of magnitude, striking a practical balance between model accuracy and characterization efficiency.

\begin{figure}[t]
  \centering

  \newcommand{\subfigheight}{3.3cm} 

  \begin{subfigure}[t]{0.49\linewidth}
    \centering
    \begin{minipage}[t][\subfigheight][c]{\linewidth}
      \centering
      \includegraphics[height=\subfigheight, width=\linewidth, keepaspectratio]{
        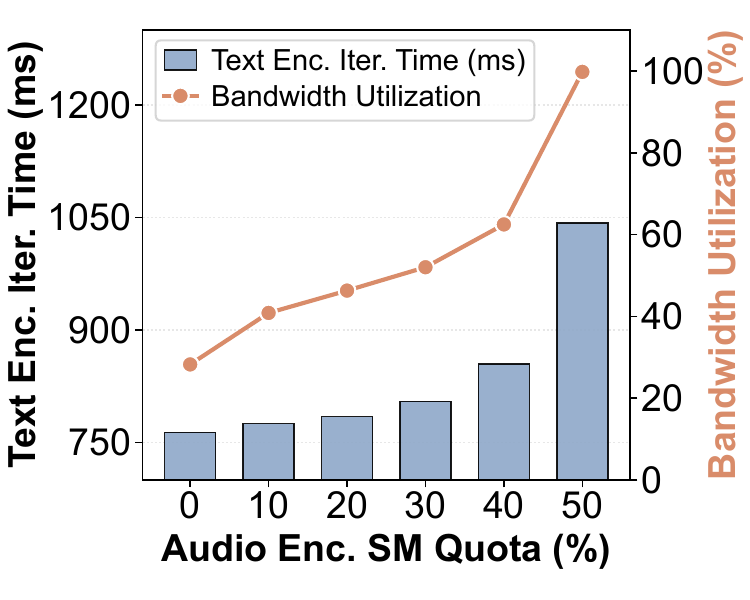
      }
    \end{minipage}
    \caption{Increasing SM quota of the audio-encoder affects efficiency of the colocated text-encoder and overall bandwidth utilization.}
    
    \label{fig:dram_overhead}
  \end{subfigure}
  \hfill
  \begin{subfigure}[t]{0.49\linewidth}
    \centering
    \begin{minipage}[t][\subfigheight][c]{\linewidth}
      \centering
      \includegraphics[height=\subfigheight, width=\linewidth, keepaspectratio]{
        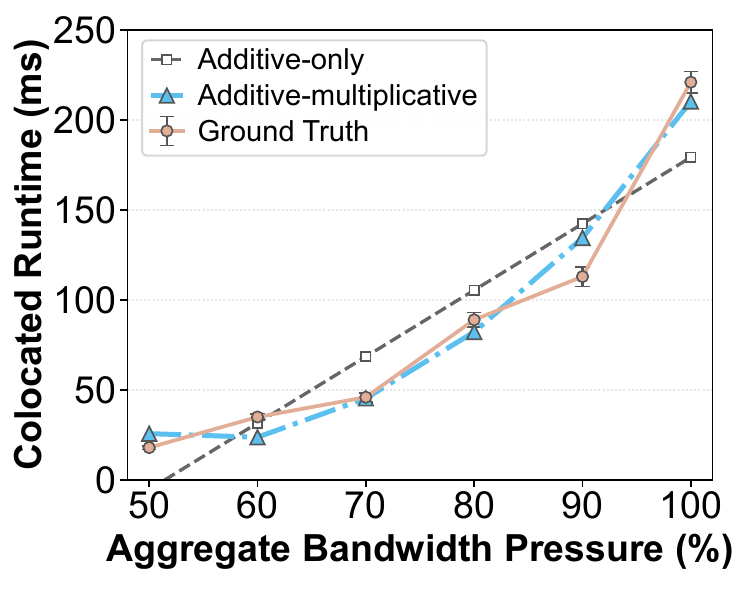
      }
    \end{minipage}
    \caption{Estimated text-encoder performance with the \emph{additive-only} and \emph{additive-multiplicative} methods, plus the ground truth.}

    \label{fig:linear_fail}
  \end{subfigure}

  \caption{Memory bandwidth contention does degrade performance, and simple linear modeling is insufficient. We train two modules (text-encoder and audio-encoder) from the OFASys model on an H100 GPU.}
  \label{fig:sm_comparison_combined}
\end{figure}

\phm{Performance model rectification with awareness to I/O interferences.}
To address the second challenge—the non-linear interference arising from co-location—\sys incorporates an I/O-aware correction mechanism into its performance model.
The scaling surface $\Tmodule(m, d_m, \alloc_m)$ constructed via grid sampling primarily captures compute-bound (SM) contention. However, MM modules are also significantly I/O-intensive. 
Under the NVIDIA GC mechanism, co-located modules share a common GPU memory plane, making memory bandwidth contention a major source of execution inefficiency. As illustrated in Fig.~\ref{fig:dram_overhead}, even with a fixed SM quota $\alloc_m$, a module's computing efficiency fluctuates remarkably as the available memory bandwidth varies. 
As in Fig.~\ref{fig:dram_overhead}, when the text-encoder's compute quota is fixed, increasing the co-located audio-encoder's compute quota aggravates bandwidth contention and further prolongs the text-encoder's iteration time.

To model this without exhaustive profiling of module colocation combinations, we use memory bandwidth utilization $B(m, \alloc_m) \in \range{0,1}$ as a generic interference proxy. This metric is collected concurrently during grid sampling at no extra cost. We then adjust the baseline latency by an I/O interference term $\Delta^\res_m$; for a module spanning multiple GPUs $\Res_m$, the execution time is governed by the worker encountering the maximum delay:
\begin{equation}
\Tmodule^{\text{rectified}}(m, \Alloc) = \Tmodule(m, \Alloc) + \max_{\res \in \Res_m} \Delta^\res_m.
\end{equation}
The remaining question is how to estimate the delay $\Delta_m^\res$ when multiple modules are co-located on the same GPU. We observe that linear approximations of aggregate utilization—used in works like LLMStation~\cite{llmstation}—fail to capture high-intensity contention. 
Specifically, as shown in Fig.~\ref{fig:linear_fail}, the additive-only model attains only $R^2=0.86$ whereas ours reaches 0.97 ($R^2$ is the \emph{coefficient of determination}~\cite{draper1998applied}, and a value closer to 1 indicates a better fitting performance), because it underestimates the rapidly increasing runtime when aggregate bandwidth pressure approaches saturation.

Consistent with recent studies~\cite{usher_osdi24, krypton_atc25}, we find that marginal interference is often multiplicatively augmented by the resource demands of peers. We therefore employ a composite formulation to fit the universal interference coefficients $e_1, e_2, e_3$:
\begin{equation}
\Delta_m^\res =  e_1 + e_2 \sum_{m \in \@M_\res} B(m, \alloc_m) + e_3 \prod_{m \in \@M_\res} B(m, \alloc_m),
\label{eq:model}
\end{equation}
where $\@M_\res = \set{m \in \stage_i | \res \in \Res_m}$ is set of modules co-located on GPU $\res$. As shown in Fig.~\ref{fig:linear_fail}, this additive-multiplicative approach accurately tracks non-linear degradation, providing the high-fidelity estimation required for optimal allocation.

\subsection{Efficient and High-quality Decision Making}
\label{subsec:mapping_solver}

\newcommand{\StageEval}[0]{\textsc{StageEval}\xspace}
\newcommand{\EarlyPrune}[0]{\textsc{EarlyPrune}\xspace}

With the ability to accurately estimate $\Tmodule(m,\Alloc)$, we can now turn to the scheduling problem as defined in Sec.~\ref{sec:problem_formulation}. \sys needs to partition modules into an ordered list of
stages $\Stage=\langle\stage_1,\stage_2,\ldots,\stage_n\rangle$, where each stage $\stage_i\subseteq\@M$ contains modules that execute in parallel, and the stage order respects every dependency in $\@E$. For each stage $\stage_i$, \sys must also choose an optimal allocation plan $\Alloc_{\stage_i}$ over the GPU set $\Res$ to minimize stage latency. The resulting iteration time is $\Titer(\Stage,\Res)=\sum_{\stage_i\in\Stage}\Tstage(\stage_i,\Res)$.

This decision has two coupled levels. The upper level decides which independent modules should be grouped into the same stage, while the lower level decides whether the modules inside a candidate stage can be placed onto the available GPUs with acceptable latency. The coupling makes exhaustive search intractable: the number of possible stage partitions grows as the Bell number of $|\@M|$, and each partition still requires solving a complex per-stage resource mapping problem. \sys address this with Alg.~\ref{alg:apollo_stage_formation} which we will detail next.



\phm{A high-quality algorithm to solve Problem~\ref{eq:stage_time}: combining GAHC heuristic with CP-SAT solver.}
\sys starts from a conservative stage list in which every module forms its own stage (Line 1). 
In each round, it considers all pairs of current stages that can be legally merged without violating the dependency order. For a candidate merge $\stage_x\cup\stage_y$, \sys invokes \StageEval to compute the optimal latency $T_{\stage_x\cup\stage_y}$ and the corresponding optimal allocation
$\Alloc_{\stage_x\cup\stage_y}$. The merge benefit is 
\begin{equation}
  \Delta(\stage_x,\stage_y)
  =
  T_{\stage_x} + T_{\stage_y}
  - T_{\stage_x \cup \stage_y}. \qquad \text{(Line 12)} 
  \label{eq:merge_saving_rewrite}
\end{equation}
\sys applies the merge with the largest positive
$\Delta$; if no merge can further reduce the iteration time, the search terminates (Line 15$\sim$18). This greedy merging is an instance of \emph{Greedy Agglomerative Hierarchical Clustering} (GAHC), a classical heuristic for progressive grouping.

\begin{algorithm}[t]
  \newcommand{\Best}[0]{\op{Best}}
  \caption{\sys's Temporal-Spatial Mapping Solver}
  \label{alg:apollo_stage_formation}
  \begin{algorithmic}[1]
    \Require MM DAG $\inner{\@M,\@E}$, GPU set $\Res$
    \Ensure Stage list $\Stage$ and allocations $\inner{\Alloc_{\stage_1}, \ldots, \Alloc_{\stage_n}}$
    \State $\Stage \gets \langle \{m\} \mid m \in \@M \rangle$ in a topological order
    \ForAll{$\stage \in \Stage$}
      \State $(T_{\stage}, \Alloc_{\stage}) \gets \StageEval(\stage,\Res)$
    \EndFor
    \While{$|\Stage| > 1$}
      \State $\Delta_{\mathrm{best}} \gets 0$; $\Best \gets \bot$
      \ForAll{$(\stage_x,\stage_y) \in \mathrm{Pairs}(\Stage)$}
        \If{merging $\stage_x$ and $\stage_y$ violates $\@E$}
          \State \textbf{continue}
        \EndIf
        \If{\EarlyPrune$(\stage_x, \stage_y, \Delta_{\text{best}})$}
            \State \textbf{continue}
        \EndIf{}
        \State $(T_{\stage_x \cup \stage_y}, \Alloc_{\stage_x \cup \stage_y})
          \gets \StageEval(\stage_x \cup \stage_y,\Res)$ \Comment{cached if seen}
        \State $\Delta \gets T_{\stage_x}+T_{\stage_y}-T_{\stage_x \cup \stage_y}$
        \If{$\Delta > \Delta_{\mathrm{best}}$}
          \State $\Delta_{\mathrm{best}} \gets \Delta$; $\Best \gets (\stage_x,\stage_y)$
        \EndIf
      \EndFor
      \If{$\Best = \bot$}
        \State \textbf{break}
      \EndIf
      \State $(\stage_x,\stage_y) \gets \Best$
      \State Apply merge of $\stage_x$ and $\stage_y$ in $\Stage$.
    \EndWhile
    \State \Return $\Stage, \inner{\Alloc_{\stage} \mid \stage \in \Stage}$
  \end{algorithmic}
\end{algorithm}

For each candidate stage proposed by the upper-level GAHC search, \StageEval solves the optimal intra-stage allocation (i.e., Eq.~\eqref{eq:stage_time}). Its min--max objective is hard to hand to a standard optimization solver as-is, so we apply binary search on a scalar target latency $\tau$ and convert it into repeated feasibility tests. For each trial $\tau$, we collect, for every module, the deployment options whose predicted latency meets the target $\tau$, then ask whether one option per module can be selected and placed within the resource constraints. This joint selection-and-placement check is devised as a multi-constraint feasibility problem and solved with established CP-SAT~\cite{ortools} engine. The smallest feasible $\tau$ is exactly $\Tstage(\stage,\Res)$.

Additionally, we set a tunable hyperparameter \textit{SM-quota search granularity} to discretize each module's fractional SM quota into candidate allocation levels. Empirically, we set it to $10\%$ and justify this choice in the Sec.~\ref{sec:eval_sensitivity}.

\phm{Enhancing algorithm efficiency with early-pruning and result-caching.}
The preceding procedure yields high-quality mappings, but it is far from fast in
practice, largely because evaluating candidate merges repeatedly invokes
\StageEval. We therefore apply \emph{early-pruning} and \emph{result-caching} to cut this cost.

First, \emph{early-pruning} (Line 9 in Alg.~\ref{alg:apollo_stage_formation}) drops candidate merges that cannot possibly beat the best merge gain $\Delta_{\mathrm{best}}$ already found in the current greedy round. Before running \StageEval, we test whether each module in the would-be
merged stage still has a \emph{non-empty} set of deployment options under which a merge gain strictly larger than $\Delta_{\mathrm{best}}$ is attainable; if any module fails, the pair cannot outperform $\Delta_{\mathrm{best}}$ and we skip the relatively costly full CP-SAT solver call.

Second, \emph{result-caching} (Line 11 in Alg.~\ref{alg:apollo_stage_formation}) memorizes the outcome of \StageEval. We store the solved latency and allocation $(T_{\stage},\Alloc_{\stage})$ the first time a stage composition $\stage$ appears and return the cached entry on later requests. This is well motivated: the outer greedy loop runs for many rounds and evaluates a large number of candidate merges per round, while committing to only \emph{one} merge at the end of each round, so the same intermediate stage compositions are revisited often---yielding plentiful cache hits and avoiding repeated CP-SAT work for identical module sets.

With the above two techniques, we can effectively reduce search overhead while preserving the solution quality, which will be confirmed by our evaluations later in Sec.~\ref{subsec:eval_superiority_mapping_solver}.



\subsection{Discussions}
\label{sec:discussion}



\textbf{Impact of data heterogeneity.}
Data heterogeneity is a hot research topic for MM training, which manifests in two aspects: \emph{intra-modal} heterogeneity and \emph{inter-modal} heterogeneity. 
We note that they do not affect the effectiveness of \sys.
Regarding intra-modal (sample-length) heterogeneity, a series of data preparation techniques, like \emph{padding}~\cite{padding}, \emph{truncation}~\cite{truncation} and \emph{pruning}~\cite{pruning}, are commonly used in mainstream training practices~\cite{internvl25,dualspeed,tokenpruning_mllm}. 
We follow such practices and ensure that the ultimate inputs of a module have the same length. 
Regarding inter-modal (sample-format) heterogeneity, we note that its impact has already been priced in when building our module-specific performance models: the module execution time is jointly affected by the module size, sample format and batch size, all of which are fixed during the training process and thereby well captured by \sys. 

\phm{Compatibility with pipeline parallelism.}
While in this paper we focus on modest-size edge-grade MMs where \emph{pipeline parallelism} (PP)~\cite{pipedream, gpipe} is inappropriate (Sec.~\ref{subsec:existing_solution}), for cloud-hosted cutting-edge MMs with hundreds of billions of parameters~\cite{qwen35_397b,gpt54,gemini31pro,kimi_k25,glm46}, PP deployment would be unavoidable, which means to host the encoders/decoders and LLM backbones in separate GPU sets---each forming a pipeline stage. 
We note that our temporal-spatial multiplexing principle can still be applied together with PP.  
With colocated module execution, we can enhance the GPU utilization of the encoder/decoder modules.
In that sense, some of their allocated GPUs can be yielded to other pipeline stages, which, with proper load balancing methods~\cite{metis,wang2026suika}, can help to further improve the end-to-end pipeline execution time.


\phm{Extensibility to inference workloads.}
When serving MM inference requests, the problem also exists that a single MM module fails to utilize the allocated GPU.
Moreover, in some scenarios, a single input sample may trigger multiple MM inference tasks: for example, in V-Bench~\cite{vbench} evaluation, 
a video sample is processed by multiple MM raters (e.g., DINO~\cite{dino}, ViT~\cite{vit} and RAFT~\cite{raft}).
In such cases, it is appealing to further enable \emph{cross-MM} module colocation to improve the overall benchmarking efficiency. 
That said, we note that the research challenges for inference workloads are different.
Different from training scenarios, in typical inference scenarios, the batch size is determined by the runtime request intensity; such batch size uncertainty yields a larger performance modeling space.
Meanwhile, the solution paradigm would also be different (e.g., batch size and resource provisioning plan may also be part of the solution).

\section{Implementation}
\label{sec:implementation}


\sys is implemented with $\sim$8K lines of code, including \textit{Mosaic Multiplexing Engine} with $\sim$3K LoC, \textit{Mosaic Performance Model} with $\sim$3K LoC, and \textit{Mosaic Mapping Solver} with $\sim$2K LoC. 
\textit{Mosaic Multiplexing Engine} is a PyTorch-based~\cite{pytorch} distributed runtime. It manages module-specific communication groups and reuses them across stages to avoid unnecessary communicator initialization. It also implements the CUDA stream pooling optimization as described in Sec.~\ref{subsec:apollo_engine}.
\textit{Mosaic Performance Model} conducts profiling for each module, sweeping chosen DP degrees and SM quotas, recording iteration time, GPU memory usage, SM utilization, and memory bandwidth utilization into structured JSON files, and further manages scaling curve fitting.
\textit{Mosaic Mapping Solver} implements Alg.~\ref{alg:apollo_stage_formation} with Google OR-Tools CP-SAT~\cite{ortools}, encoding group selection, GPU placement, and SM-quota assignment as integer variables with memory under SM-capacity constraints. It stores colocation groups as module bitmasks to realize result-caching for \StageEval, and also applies early-pruning for merging candidates before calling the solver.


\section{Evaluation}
\label{sec:evaluation}



This section evaluates \sys performance on representative MM training workloads over state-of-the-art baseline methods (Sec.~\ref{sec:eval_setup}).
The end-to-end performance results in Sec.~\ref{eval:end-to-end} suggest that \sys can remarkably improve the training efficiency and GPU utilization.
Further microscopic studies in Sec.~\ref{sec:eval_micro} demonstrate the effectiveness of each of \sys's core design innovations.
In Sec.~\ref{sec:eval_sensitivity} we conduct sensitivity studies to evaluate \sys performance across diverse resource scales and SM-quota search granularity.

\subsection{Experimental Setup}
\label{sec:eval_setup}

\phm{Hardware.}
All experiments are conducted on a GPU cluster consisting of 4 HGX nodes, each equipped with 8 NVIDIA H100 (80GB) GPUs and 64 Intel(R) Xeon(R) Platinum 8468 CPU cores. 
The CUDA version is 13.0.
Within each node, GPUs are connected by NVLink with 900 GBps bandwidth. 
Across nodes, RDMA communication is provided by 8 InfiniBand NICs per node, each with 400 Gbps bandwidth.


\phm{Models.}
In our experiments, we respectively train six representative MM models: \textit{CLIP}~\cite{clip}, \textit{Qwen3-VL}~\cite{qwen3vl}, \textit{Unified-IO~2}~\cite{unifiedio2}, \textit{ImageBind}~\cite{imagebind}, \textit{OFASys}~\cite{ofasys}, and \textit{CTVLM}~\cite{ctvlm}.
All of them are elaborated in Sec.~\ref{sec:research_background}.
In particular, \textit{Unified-IO~2}, \textit{ImageBind}, and \textit{OFASys} each can be viewed as a model \emph{family}, in which, depending on the multi-modal tasks, the modules provisioned can be flexibly combined to form diverse MMs.
Regarding the training input data, as shown in Table~\ref{tab:mockdata_config}, we apply the data preparation methods discussed in Sec.~\ref{sec:discussion} to enforce a uniform (yet modal-specific) sample length (size) for each module; all are typical values suggested by their official documentations. 
The training batch size is by default set to 32 and degrades to 16 for single-node experiments. 


\begin{table}[t]
\centering
\small
\setlength{\tabcolsep}{6pt}
\renewcommand{\arraystretch}{0.88}
\begin{tabular}{p{0.20\linewidth}p{0.34\linewidth}p{0.30\linewidth}}
\toprule
\textbf{Modality} & \textbf{Parameter} & \textbf{Value} \\
\midrule
Text & Sequence length & 2048 tokens \\
\midrule
\multirow{2}{*}{Image} & Resolution & $512{\times}512$ \\
& Channels & RGB \\
\midrule
\multirow{2}{*}{Video} & Frames & 32 \\
& Resolution & $512{\times}512$ \\
\midrule
\multirow{2}{*}{Audio} & Sampling rate & 16 kHz \\
& Duration & 8 s \\
\midrule
{Depth} & Resolution & $224{\times}224$ \\
\midrule
{Thermal} & Resolution & $256{\times}256$ \\
\midrule
\multirow{3}{*}{IMU} & Axes & 6 \\
& Sampling rate & 100 Hz \\
& Duration & 8 s \\
\midrule
Action & Sequence length & 256 tokens \\
\midrule
Box & Coordinates & $(x_1,y_1,x_2,y_2)$ \\

\bottomrule
\end{tabular}
\caption{Training data configurations.}
\label{tab:mockdata_config}
\end{table}

\phm{Baselines and metrics.}
We compare \sys with the three baselines previously elaborated in Sec.~\ref{subsec:existing_solution}: \emph{Megatron-LM}~\cite{megatron}, \emph{DistMM}~\cite{distmm}, and \emph{Spindle}~\cite{spindle}. 
Our primary metric to evaluate the training efficiency is the \emph{per-iteration training makespan}. 
Additionally, we report hardware efficiency via \emph{Compute-Warps-in-Flight}~\cite{gpu_metrics}\footnote{We avoid the misleading coarse-grained GPU Utilization as it only reflects whether any kernel is active and may remain high even when SM resources are under-saturated. In contrast, Compute-Warps-in-Flight directly reflects the amount of compute-active warps on SMs, making it a better indicator.}.

\subsection{End-to-End Performance}
\label{eval:end-to-end}

\begin{figure}[t]
    \centering
    \includegraphics[width=0.9\linewidth]{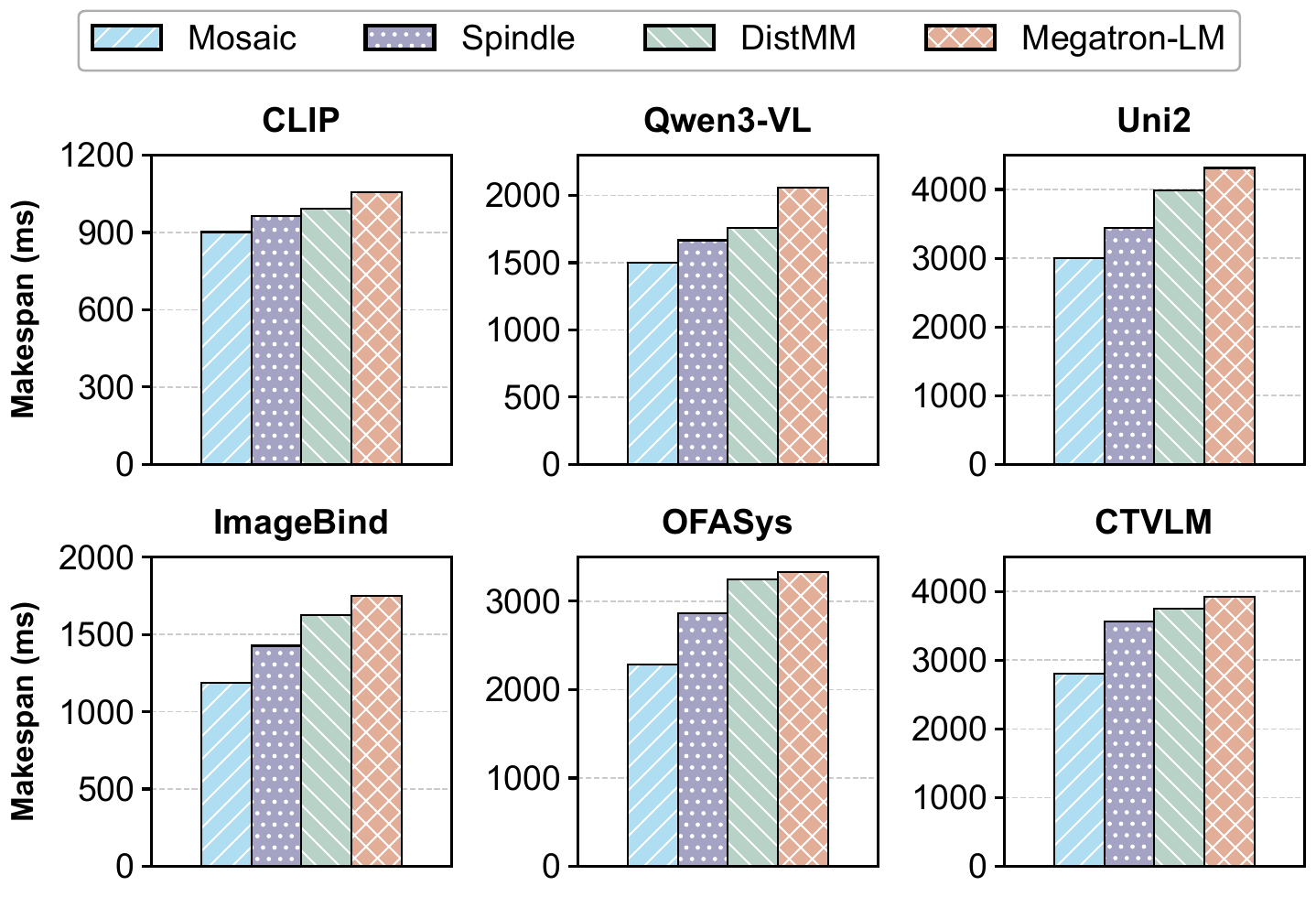}
    \caption{Per-iteration time when training different MMs.} 
    \label{fig:lmm_makespan_grid}
\end{figure}

\begin{figure}[t]
    \centering
    \includegraphics[width=0.9\linewidth]{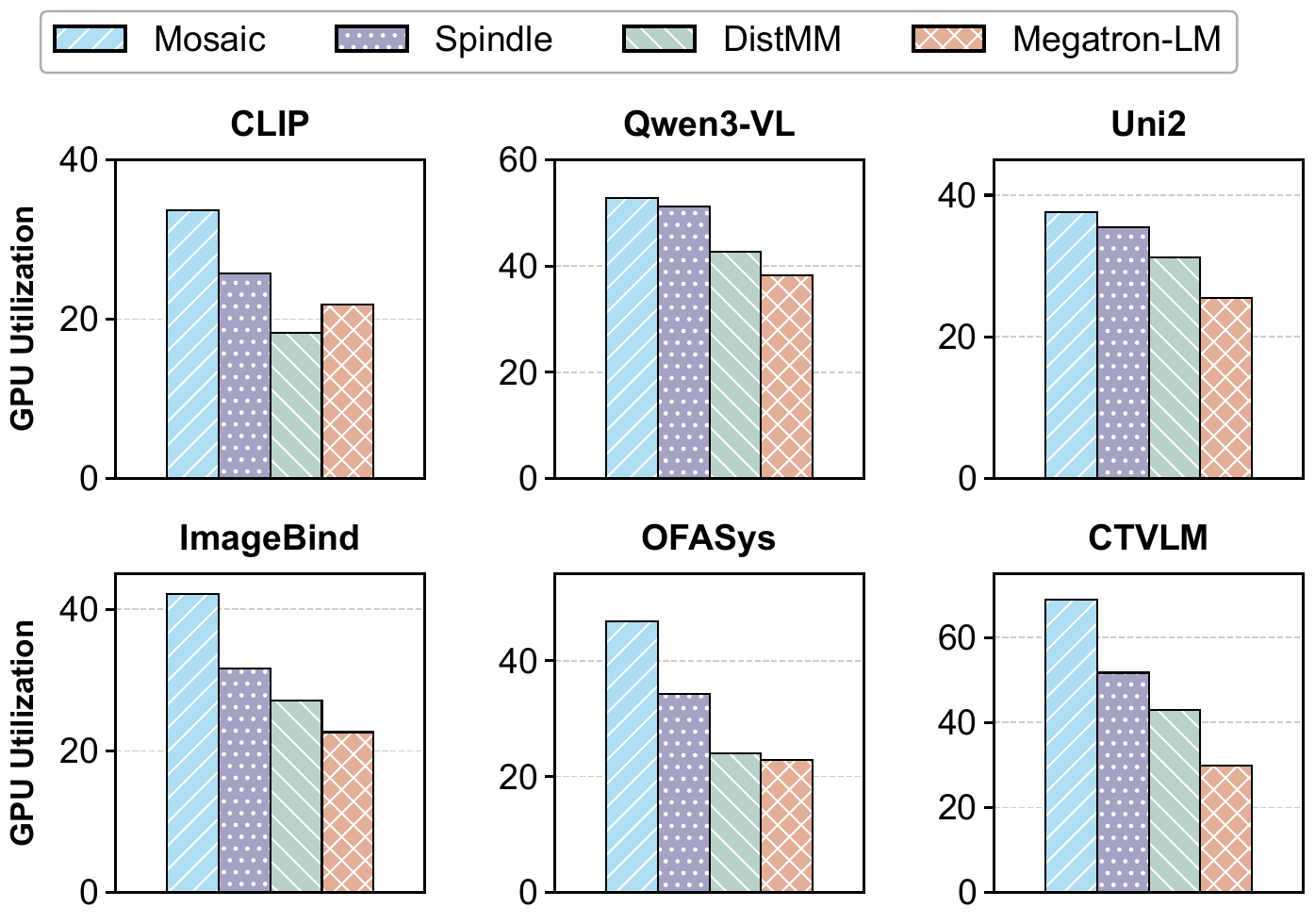}
    \caption{GPU utilization when training different MMs.}
    \label{fig:warp_in_flight}
\end{figure}

Fig.~\ref{fig:lmm_makespan_grid} depicts the average per-iteration time of each MM under different deployment methods. 
It confirms that \sys consistently achieves the best training efficiency, with $1.07\times$--$1.31\times$ speedup over \textit{Spindle} (the second best), $1.10\times$--$1.42\times$ over \textit{DistMM}, and $1.17\times$--$1.48\times$ over \textit{Megatron-LM}.
Such superiority aligns with our previous analysis in Sec.~\ref{subsec:existing_solution}. 
Moreover, we also notice that the performance benefit of \sys is relatively larger for more complex MMs: the improvement of \sys over \emph{Spindle} is {$1.07\times$} for CLIP, yet it is {$1.31\times$} for OFASys (which has 9 encoder modules).
This is because for MMs with substantially heterogeneous modules, enforcing full-GPU allocation, as done in existing works, would render some GPUs highly underutilized.  


To confirm, we further resort to the average per-GPU utilization information shown in Fig.~\ref{fig:warp_in_flight}.
It suggests that \sys does achieve the highest GPU utilization for all the models: on average, the GPU utilization under \sys is 47.0\%, yet under \textit{Spindle}, \textit{DistMM}, and \textit{Megatron-LM}, they are respectively 38.3\%, 31.0\%, and 26.8\%. 
In particular, for the highly-heterogeneous OFASys model, the most underutilized GPU under \textit{Spindle}, which hosts the \emph{IMU} module, has a utilization of {$24.7\%$}, yet under \sys that utilization is increased to {$56.3\%$}.
This demonstrates that, by enabling spatial multiplexing, \sys effectively enhances the overall GPU utilization, thereby improving the MM training efficiency. 


\subsection{Microscopic Performance Deep Dive}
\label{sec:eval_micro}

In this part, we respectively study the effectiveness of a series of techniques adopted in \sys.

\begin{figure}[t]
    \centering
    
    \begin{subfigure}[t]{0.49\linewidth}
        \centering
        \vspace{0pt}
        \includegraphics[width=\linewidth]{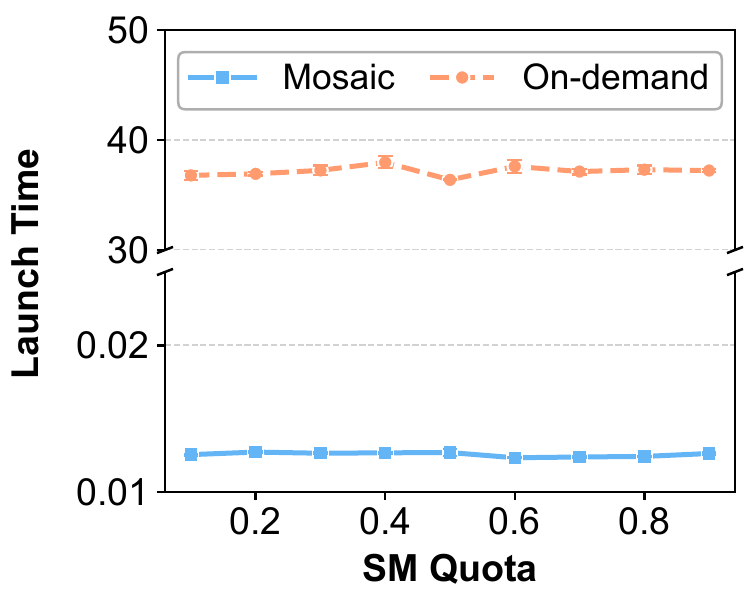}
        \caption{GC-stream preparation latency under different schemes.}
        \label{fig:gctx_controller_overhead}
    \end{subfigure}
    \hfill
    \begin{subfigure}[t]{0.49\linewidth}
        \centering
        \vspace{0pt}
        \includegraphics[width=\linewidth]{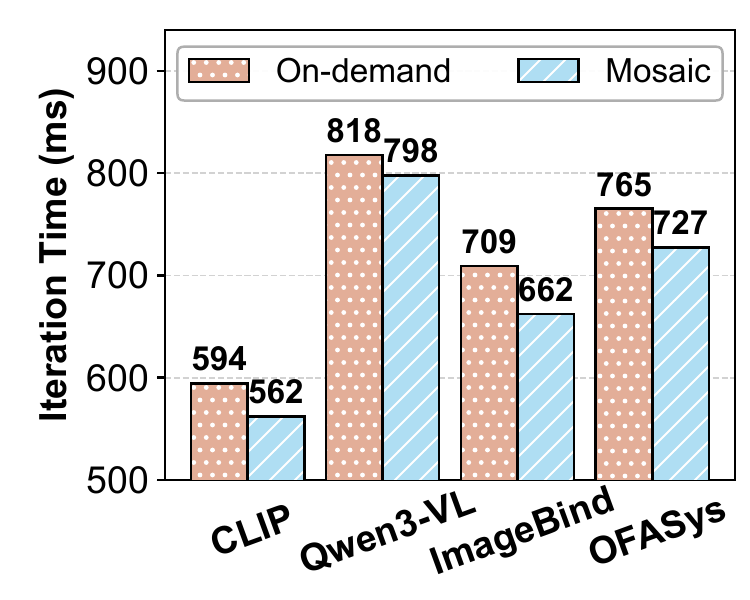}
        \caption{End-to-end iteration time under different schemes.}
        \label{fig:gctx_controller_speedup}
    \end{subfigure}

    \caption{
        GC-stream pre-creation remarkably reduces provision latency and benefits end-to-end iteration time.
    }
    \label{fig:gctx_controller}
\end{figure}

\subsubsection{Superiority of \sys Multiplexing Engine.} 
Recall that in Sec.~\ref{subsec:apollo_engine}, to mitigate the runtime provisioning overhead of diverse GC-streams, we propose to pre-create a pool of GC-streams with the desired quotas. 
To check its necessity, we measure the average time overhead to create GC-stream of different SM quotas (from 0.1 to 0.9; each point is repeated 30 times, and the error-bars show the upper/lower bounds).
As shown in Fig.~\ref{fig:gctx_controller_overhead}, on-demand GC-stream creation takes a time overhead of around 37 ms; in contrast, by creating the GC-streams a priori, the amortized GC-stream launching overhead shrinks to around 0.013 ms, achieving a remarkable overhead reduction. 

We further evaluate how this GC-stream launching speedup translates to the end-to-end performance benefit.
Fig.~\ref{fig:gctx_controller_speedup} compares the average per-iteration time of four typical MMs (CLIP, Qwen3-VL, ImageBind, and OFASys) with and without GC-stream pre-creation. 
Across these workloads, \sys reduces end-to-end iteration time by $2.4\%$--$6.6\%$, with an average reduction of about $4.6\%$.
This confirms that such a technique is indispensable to \sys performance superiority.	

\subsubsection{Superiority of \sys Performance Model.}

\begin{figure}[t]
    \centering
    \begin{subfigure}[t]{0.48\linewidth}
        \centering
        \includegraphics[width=\linewidth]{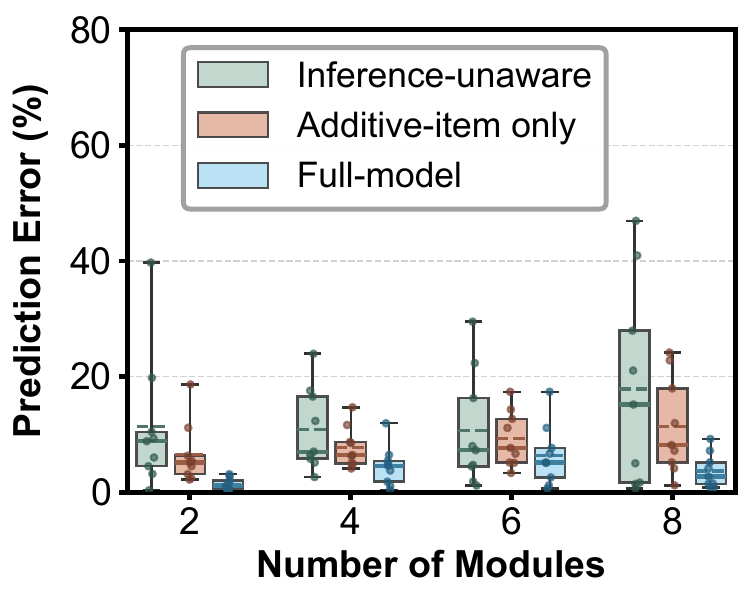}
        \caption{Prediction error across different modeling strategies.}
        \label{fig:prediction_error}
    \end{subfigure}
    \hfill
    \begin{subfigure}[t]{0.48\linewidth}
        \centering
        \includegraphics[width=\linewidth]{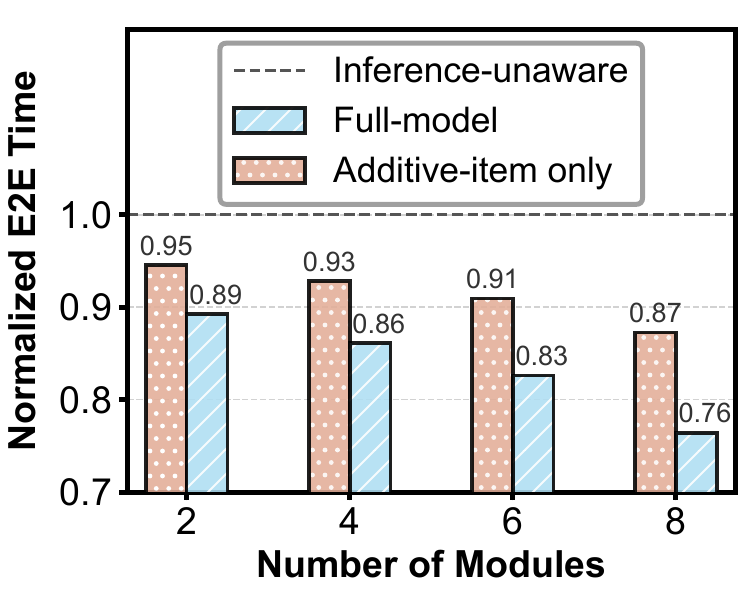}
        \caption{Normalized E2E time under different modeling strategies.}
        \label{fig:e2e_time_with_diff_predict}
    \end{subfigure}

    \caption{\sys's interference-aware performance model outperforms the other baseline methods in both prediction error and end-to-end performance.}
    \label{fig:prediction_error_and_e2e}
\end{figure}

Recall that in Sec.~\ref{subsec:performance_modeling}, we have built an interference-aware performance model which includes both the \emph{additive} and \emph{multiplicative} items.
To confirm the superiority of that performance model, we resort to the ablation study depicted in Fig.~\ref{fig:prediction_error}, in which we evaluate the modeling accuracy for the OFASys model by setting up varying module numbers. 
Specifically, we evaluate three modeling schemes: (1) \emph{inference-unaware}, which directly uses the per-module scaling surfaces, (2) \emph{additive-item only}, which only includes the additive item when fitting the performance model, and (3) \emph{full model}, which uses the full model depicted by Eq.~\ref{eq:model}. 
{The ground-truth latency is obtained by actually executing each tested colocation plan.}
As shown in Fig.~\ref{fig:prediction_error}, the \emph{full-model} method consistently attains the highest modeling accuracy. 
For example, with 8 modules in the {OFASys} model, it attains an average prediction error of merely {3.65\%}, 
much smaller than the other methods (11.39\% and 17.85\%, respectively). 
This confirms the superiority of our performance model in Eq.~\ref{eq:model}.

In Fig.~\ref{fig:e2e_time_with_diff_predict}, we further show the resultant per-iteration time of the {OFASys} model under different performance modeling methods; all results are normalized by the value under the \emph{interference-unaware} performance model.
Fig.~\ref{fig:e2e_time_with_diff_predict} reveals that the performance model in Eq.~\ref{eq:model} achieves a training speedup of 11\%--24\%.
Moreover, that speedup level increases when the {OFASys} model comprises more modules, because with more modules, the cross-module performance interference is more salient. 
In summary, our interference-aware performance model is crucial to the superiority of \sys.

\begin{figure}[t]
    \centering

    \begin{subfigure}[t]{0.48\linewidth}
        \centering
        \includegraphics[width=\linewidth]{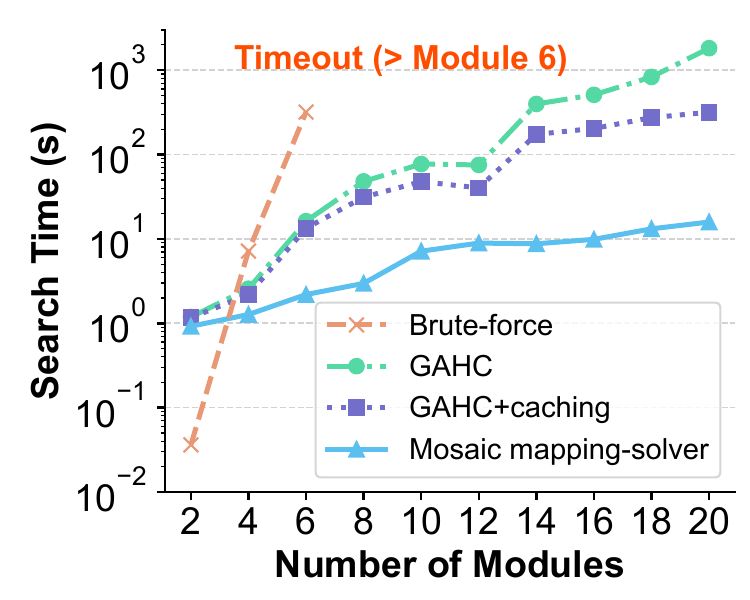}
        \caption{Search time ablation study.}
        \label{fig:allocation_search_comparison}
    \end{subfigure}
    \hfill
    \begin{subfigure}[t]{0.48\linewidth}
        \centering
        \includegraphics[width=\linewidth]{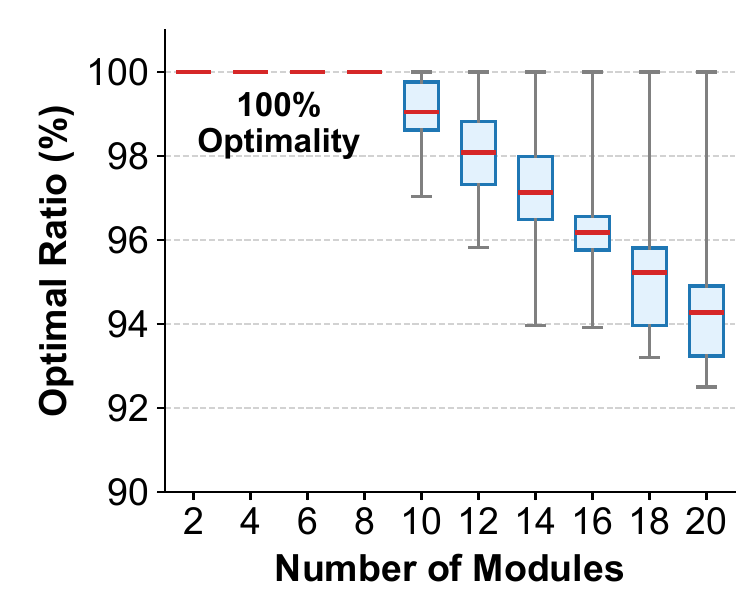}
        \caption{Quality of \sys's solution compared to genuine optimal from exhaustive enumeration.}
        \label{fig:search_optimal_solution}
    \end{subfigure}

    \caption{\sys's search algorithm finds near-optimal solution with high time efficiency.}
    \label{fig:allocation_search_and_optimality}
\end{figure}

\subsubsection{Superiority of \sys Mapping Solver.}
\label{subsec:eval_superiority_mapping_solver}

Recall that in Sec.~\ref{subsec:mapping_solver}, to solve the \emph{module-stage} mapping and \emph{module-GPU} mapping problem, we apply the GAHC heuristic and also incorporate \emph{early-pruning} and \emph{result-caching} for better efficiency.
To confirm the effectiveness of those methods, we compare the efficiency and quality performance of \sys mapping-solver (Alg.~\ref{alg:apollo_stage_formation}) against multiple baselines: (1) \textit{brute-force}, which finds the best mapping scheme by exhaustive grid search, (2) \textit{GAHC}, which uses the standard GAHC algorithm without pruning or caching, (3) \textit{GAHC+caching}, which further incorporates result caching into {GAHC}, and (4) \textit{\sys mapping-solver}, which incorporates both early-pruning and result-caching with GAHC.

Fig.~\ref{fig:allocation_search_comparison} shows the search time of the optimal deployment plan for the OFASys model (under varying module number).
According to Fig.~\ref{fig:allocation_search_comparison}, as the module number increases, the \textit{brute-force} method quickly becomes intractable (surpassing the time budget of 1800s).
By contrast, all the other methods can find the optimal solution within the time budget for up to 20 modules. 
In particular, the \textit{\sys mapping-solver}, which combines GAHC with both early-pruning and result-caching, attains the best efficiency; this demonstrates the necessity of the two techniques.

Meanwhile, regarding the solution quality, as suggested in Fig.~\ref{fig:search_optimal_solution} (the optimal plan is obtained by exhaustive search), the deployment plan found by our \sys mapping-solver achieves $100\%$ optimality (in per-iteration time) for up to 4 modules, and remains near-optimal as the search space grows.
The median optimality ratio is around $94.27\%$ with 10 modules.
Therefore, \sys mapping-solver performs well in both solution efficiency and quality.



\subsection{Sensitivity Analysis}
\label{sec:eval_sensitivity}


\subsubsection{Sensitivity to GPU Cluster Scale.}

\begin{figure}[t]
    \centering
    \begin{subfigure}[t]{0.48\linewidth}
        \centering
        \includegraphics[width=\linewidth]{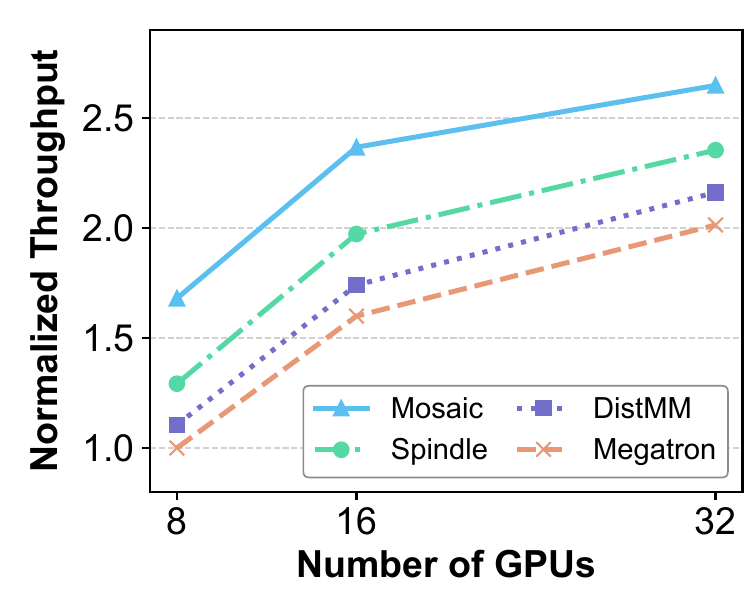}
        \caption{Throughput under different GPU cluster scales.}
        \label{fig:throughput_with_increasing_gpu}
    \end{subfigure}
    \hfill
    \begin{subfigure}[t]{0.48\linewidth}
        \centering
        \includegraphics[width=\linewidth]{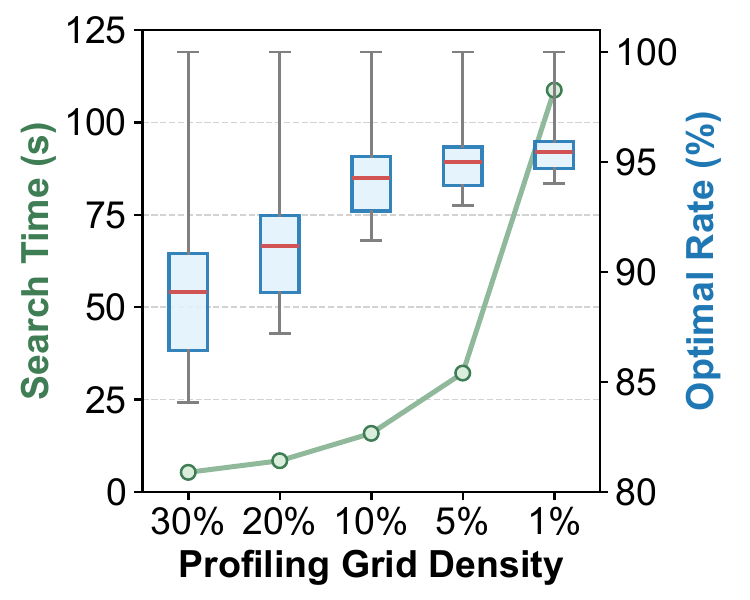}
        \caption{Search-time/quality trade-off under different SM-quota search granularities.}
        \label{fig:sm_quota_granularity_tradeoff}
    \end{subfigure}
    \caption{Sensitivity of \sys to cluster scale and SM-quota search granularity on the OFASys model.}
    \label{fig:sensitivity_analysis}
\end{figure}

We first examine whether \sys remains effective as the available resource pool scales.
This experiment fixes a four-module \textit{OFASys} workload, varies the GPU pool size from 8 to 32 GPUs and compares their throughput (defined as $\frac{1}{\text{iteration time}}$).
As shown in Fig.~\ref{fig:throughput_with_increasing_gpu}, \sys consistently achieves the highest throughput across all evaluated GPU pool sizes.
Compared with \textit{Megatron-LM} under the same GPU pool size, \sys improves throughput by \(1.68\times\), \(1.48\times\), and \(1.31\times\) on 8, 16, and 32 GPUs, respectively.
Against the strongest baseline, \textit{Spindle}, the corresponding speedups are \(1.30\times\), \(1.20\times\), and \(1.12\times\); against \textit{DistMM}, they are \(1.52\times\), \(1.36\times\), and \(1.22\times\).

This trend indicates that spatial multiplexing is particularly valuable when the GPU pool is tight.
With fewer GPUs, integer and exclusive-GPU allocation policies have less room to absorb module-level imbalance, so lightweight modules are more likely to leave sub-GPU compute capacity unused.
\sys instead allows complementary modules to share GPUs with fractional SM quotas, improving aggregate throughput under the same resource budget.
As the GPU pool grows, the relative pressure on each allocation decision decreases, and the speedup gradually narrows, but \sys still preserves the best throughput at every scale.

\subsubsection{Sensitivity to SM-quota Search Granularity.}

We next study the effect of the \textit{SM-quota search granularity} introduced in Sec.~\ref{subsec:mapping_solver}, to understand how \sys balances search time and solution quality.

Fig.~\ref{fig:sm_quota_granularity_tradeoff} reports the resulting search time and median optimality ratio compared to the optimal plan found by exhaustive enumeration.
Coarse granularities substantially reduce search overhead but compromise the solution quality: at 30\% and 20\%, the search finishes in \(5.32\,\mathrm{s}\) and \(8.46\,\mathrm{s}\), while the median optimality ratio is only \(89.10\%\) and \(91.18\%\).
Reducing the granularity to 10\% (the default value for main experiments) improves the median optimality ratio to \(94.27\%\) with a moderate \(15.87\,\mathrm{s}\) search time, forming a practical knee point in the trade-off.
Further refining the search granularity brings limited additional benefit.
Compared with 10\%, the 5\% and 1\% settings make the search \(2.03\times\) and \(6.85\times\) slower, respectively, but improve median optimality by only 0.73 and 1.18 percentage points.
These results suggest that a 10\% SM-quota search granularity captures most of the benefit of fine-grained spatial allocation while avoiding the rapidly increasing overhead of near-continuous search.

\section{Related Works}
\label{sec:related}

In this section, we present other related works additional to those we have mentioned in Sec.~\ref{subsec:existing_solution}, in aspects respectively in distributed model deployment, spatial GPU multiplexing and efficient training of large-scale MMs.

\phm{Distributed model deployment.} 
With the increasing use of neural network models, deploying large neural network models on distributed GPUs has been a hot research topic in the literature.
Apart from the traditional DP~\cite{dp}, TP~\cite{megatron} and PP~\cite{pipedream,gpipe} parallelization schemes which are elaborated in Sec.~\ref{sec:research_background}, recently some new parallelization schemes also emerge, including \emph{Sequence Parallel} (SP)~\cite{sp}, \emph{Context Parallel} (CP)~\cite{cp} and \emph{Expert Parallel} (EP)~\cite{ep}.
Meanwhile, given the increasing model deployment complexity, a series of auto-parallelization works, like \emph{Alpa}~\cite{alpa} and \emph{Metis}~\cite{metis} have also been proposed. 
However, these works commonly focus on Large Language Models, which are highly symmetric and executed in a purely sequential manner; their parallelization methods are over-simplified for MMs that have multiple heterogeneous module branches.

\phm{Spatial GPU Multiplexing.}
Spatial GPU multiplexing has been widely studied to improve GPU utilization, overwhelmingly for inference acceleration.
For example, \textit{Salus}~\cite{salus} provides fine-grained GPU sharing primitives for DL applications, \textit{REEF}~\cite{reef} enables microsecond-scale preemption for concurrent DNN inferences, \textit{Orion}~\cite{orion} further improves GPU sharing performance for ML applications through inter-ference-aware scheduling, and \textit{MuxServe}~\cite{muxserve} exploits spatial-temporal multiplexing for LLM serving. 
To the best of our knowledge, we are the first that exploit spatial GPU multiplexing to speed up the training efficiency of emerging MMs.

\phm{Efficient training of large-scale, cloud-grade MMs.}
Deploying large-scale, cloud-grade MMs in hundreds or thousands of GPUs has been the research focus of a series of recent works. 
For example, \textit{Cornstarch}~\cite{cornstarch}, \textit{Optimus}~\cite{optimus_atc25}, \textit{GraphPipe}~\cite{graphpipe}, \textit{PipeWeaver}~\cite{pipeweaver}, and \textit{MegaScale-Omni}~\cite{megascale_omni} all seek to advance the parallelization schemes for such large-scale MMs, for which PP optimization (against module and data heterogeneity) is the key focus.
By contrast, in this paper, we focus on \emph{edge-grade} MMs, which would be even more significant than cloud-grade MMs in terms of the entity number and scenario diversity~\cite{edge_ai_market}. 
Meanwhile, as elaborated in Sec.~\ref{sec:discussion}, our temporal-spatial multiplexing method can be jointly applied with more advanced parallelization schemes (exemplified by PP) when training large MMs.

\section{Conclusion}
\label{sec:conclusion}

This paper introduced \sys, a system for efficient training of MMs comprising heterogeneous modules. 
Motivated by the growing module heterogeneity of modern multimodal models, 
\sys maps dependent modules into sequential stages while allowing parallel modules within each stage to share GPUs with fine-grained SM allocations. 
\sys realizes this approach with three coordinated components: a flexible GreenContext-based multiplexing engine, a performance model that captures scaling behavior and colocation interference, and a mapping solver that efficiently finds high-quality temporal and spatial deployment plans. Our evaluation on representative MMs and a 32-H100 cluster shows that \sys consistently improves GPU utilization, achieving up to 1.31$\times$ training-efficiency improvement over existing methods. 
Overall, \sys shows that carefully planned temporal-spatial multiplexing can effectively reduce cross-module bubbles and unlock underutilized GPU resources in MM training.

\bibliographystyle{acm}
\bibliography{main}

\end{document}